\documentclass[5p]{elsarticle}

\usepackage{hyperref}
\usepackage{lineno,hyperref}
\modulolinenumbers[5]
\usepackage{amssymb}
\usepackage{amsmath}
\usepackage{float}
\biboptions{sort&compress}

\journal{Acta Materialia}
\begin{document}

\newcommand{\qgb}{q_{\text{GB}}}
\newcommand{\nuo}{\overline{\nu}}
\newcommand{\os}{\overline{s}}
\newcommand{\bigcdot}{\scalerel*{\cdot}{\bigodot}} 
\newcommand{\sbullet}{\scalerel*{ \bullet}{o}}
\newcommand{\dotr}{\mbox{$\boldsymbol{\cdot}$}}

\newcommand{\Dref}{D_{\text{ref}}}
\newcommand{\cref}{c_{\text{ref}}}

\newcommand{\VAl}{\operatorname{V_{Al}^{3-}}}
\newcommand{\VO}{\operatorname{V_{O}^{2+}}}
\newcommand{\hdot}{\operatorname{ h^{+} }}
\newcommand{\eprime}{\operatorname{e^{-}}}
\newcommand{\OO}{\operatorname{O_O^{\times}}}
\newcommand{\Otwo}{\operatorname{O_2}}
\newcommand{\Al}{\operatorname{Al_{Al}^{\times}}}

\newcommand{\alumina}{\operatorname{Al_2O_3}} 

\newcommand{\Ph}{P_{\text{O}_2}^{\text{hi}}}
\newcommand{\PIeq}{P_{I}^{\text{eq}}}
\newcommand{\Pl}{P_{\text{O}_2}^{\text{lo}}}
\newcommand{\PI}{P_{\text{O}_2}^{I}}
\newcommand{\PII}{P_{\text{O}_2}^{II}}
\newcommand{\POtwo}{P_{\text{O}_2}}
\newcommand{\Peq}{P_{\text{O}_2}^{\text{eq}}}
\newcommand{\Ivac}{I_{\text{vac}}}
\newcommand{\Iion}{I_{\text{ion}}}
\newcommand{\Iel}{I_{\text{el}}}
\newcommand{\Inet}{I_{\text{net}}}
\newcommand{\kBT}{k_{\text{B}}T}
\newcommand{\lD}{\ell_{\text{D}}}
\newcommand{\LD}{\lambda_{\mathrm{D}} }

\newcommand{\sub}{s} 
\newcommand{\ozO}{\bar{z}_{\text{O}}}
\newcommand{\zO}{z_{\text{O}}}
\newcommand{\zVO}{z_{\VO}}
\newcommand{\ozAl}{\bar{z}_{\text{Al}}}
\newcommand{\zAl}{z_{\text{Al}}}
\newcommand{\zVAl}{z_{\VAl}}
\newcommand{\oz}{\bar{z}_{\text{O}}}
\newcommand{\ozsub}{\bar{z}_{\sub}}
\newcommand{\zsub}{z_{\sub}}

\newcommand{\JO}{J_{\text{O}}}
\newcommand{\JVO}{J_{\VO}}
\newcommand{\JAl}{J_{\text{Al}}}
\newcommand{\JVAl}{J_{\VAl}}
\newcommand{\muszs}{\mu_{s}^{z_s}}
\newcommand{\muOz}{\mu_{\text{O}}^{\zO}}
\newcommand{\muVO}{\mu_{\VO}}
\newcommand{\muO}{\mu_{\text{O}}}
\newcommand{\muOtwo}{\mu_{\text{O}_2}}
\newcommand{\muAl}{\mu_{\text{Al}}}

\newcommand{\irrzone}{\mathbb{IZ}}
\newcommand{\AGB}{\mathbb{A}_{\text{GB}}}
\newcommand{\ahex}{a_{\text{hex}}}

\newcommand{\sgn}{\text{sgn}}

\newcommand{\Lsp}{L_{s s}'}

\newcommand{\mmps}{\text{m}^2\,\text{s}^{-1}}


\begin{frontmatter}

\title{A model for time-dependent grain boundary diffusion of ions and electrons through a film or scale,
with an application to alumina}

 \author[a]{M.P. Tautschnig\corref{corrauthor}}
 \cortext[corrauthor]{Corresponding author.}
 \ead{mpt13@imperial.ac.uk}

 \author[b]{N.M. Harrison}

 \author[a,c]{M.W. Finnis}

 \address[a]{Department of Physics, Imperial College London, London SW7 2AZ, UK}
 \address[b]{Department of Chemistry, Imperial College London, London SW7 2AZ, UK}
 \address[c]{Department of Materials, Imperial College London, London SW7 2AZ, UK}

\begin{abstract}
A model for ionic and electronic grain boundary
transport through thin films, scales or membranes with columnar grain structure is introduced.
The grain structure is idealized as a lattice of identical hexagonal cells -- a honeycomb pattern.
Reactions with the environment constitute the boundary conditions and drive the transport between the surfaces.
Time-dependent simulations solving the Poisson equation 
self-consistently with the Nernst-Planck flux
equations for the mobile species are performed. In the resulting Poisson-Nernst-Planck 
system of equations, the electrostatic potential is obtained from the Poisson equation in its integral form 
by summation.
The model is used to interpret alumina membrane oxygen permeation experiments, in which different oxygen gas pressures
are applied at opposite membrane surfaces and the resulting flux of oxygen molecules through the membrane is measured. 
Simulation results involving four mobile species, 
charged aluminum and oxygen vacancies, electrons, and holes, provide a complete description of the measurements and 
insight into the microscopic processes underpinning the oxygen permeation of the membrane. 
Most notably, the hypothesized transition between $p$-type and $n$-type ionic conductivity of 
the alumina grain boundaries as a function of the applied oxygen gas pressure is observed in the simulations. 
The range of validity of a simple analytic model for the oxygen permeation rate, 
similar to the Wagner theory of metal oxidation, 
is quantified by comparison to the numeric simulations.
The three-dimensional model we develop here is readily adaptable to problems such as transport in a solid state 
electrode, or corrosion scale growth.
\end{abstract}


\end{frontmatter}


\section{Introduction}
Thin films of insulating material at metal -- gas and metal -- liquid interfaces accomplish a range of 
service functions in materials technology. Common examples are 
functional ceramics in electronics, energy related applications and sensors.
Thin  films formed by surface oxidation of a metal
can have either beneficial or corrosive effects. 
Alumina and chromia formed by thermal oxidation are examples of protective oxide films, 
which find application in thermal barrier coatings, 
and can be engineered for durability by additions of rare earth elements~\cite{Evans2001,Padture2002a,Stott1995,Naumenko2016}.
They can also grow in an uncontrolled manner, adhering weakly to the metal and allowing corrosion to proceed. 
Metal oxidation is a heterogeneous process, 
consisting of multiple steps, involving
the dissociation of molecular oxygen, 
transport through the growing oxide layer, 
and the reaction between oxygen and metal atoms. 
Generally speaking a growing oxide layer on the metal surface requires either metal atoms, normally ions, to 
be transported to the oxide -- gas interface to sustain the oxidation reaction, 
or alternatively oxygen atoms or ions to be transported through 
the oxide to the metal -- oxide interface to sustain an internal oxidation reaction. 
Furthermore, the transport is usually thought to be effected by diffusion of cation or anion vacancies.
Both processes may proceed, depending on the material under consideration
and on the environmental conditions such as temperature and oxygen partial pressure, $\POtwo$. 
While the slowest process is rate-determining for consecutive processes, 
like dissociation of oxygen molecules and their transport through
the oxide, the fastest process is rate-determining for parallel processes, like bulk and grain boundary diffusion 
through the oxide.
The measured oxygen and aluminum diffusion coefficients in $\alpha$-alumina
are found to be several orders of magnitude greater
at the grain boundaries than in the bulk material~\cite{Heuer2013, Heuer2008, Heuer2016}.
The fact that grain boundaries provide the dominant transport mechanism 
underlines the importance of including their geometric 
and transport properties in a realistic model of the process. 
Since vacancies in strongly ionic oxides such as 
alumina or chromia are charged species relative to the perfect crystal, 
the fluxes of these species carry an electric current, 
which in the usual scenario of steady-state growth is not sustainable, 
unless compensated by an equal and opposite current of electrons or holes, 
as described by the classic model of Wagner~\cite{Wagner1933,Atkinson1985}.
The prediction of the growth behaviour of thin films, 
and its influence on the material or device performance, requires us to describe 
the mixed ionic, electronic transport through the films, while taking their grain boundary structure into account.\\
\indent Because of the widespread importance of alumina films \cite{Dorre1984}, 
and since it is a relatively well characterized material,
we focus on alumina films for the validation of our modelling approach.
Moreover, a recent series of permeation experiments for $\alpha$-alumina polycrystal membranes, e.g.~\cite{Kitaoka2009},
conducted for different combinations 
of applied oxygen gas pressures at high temperatures,
provides an ideal test case for our transport model.
These experimental results will be summarized briefly in the following section.

\subsection{Brief review of oxygen permeation and diffusion experiments}
\begin{figure}[H]
 \begin{center}
     \includegraphics[width=5cm]{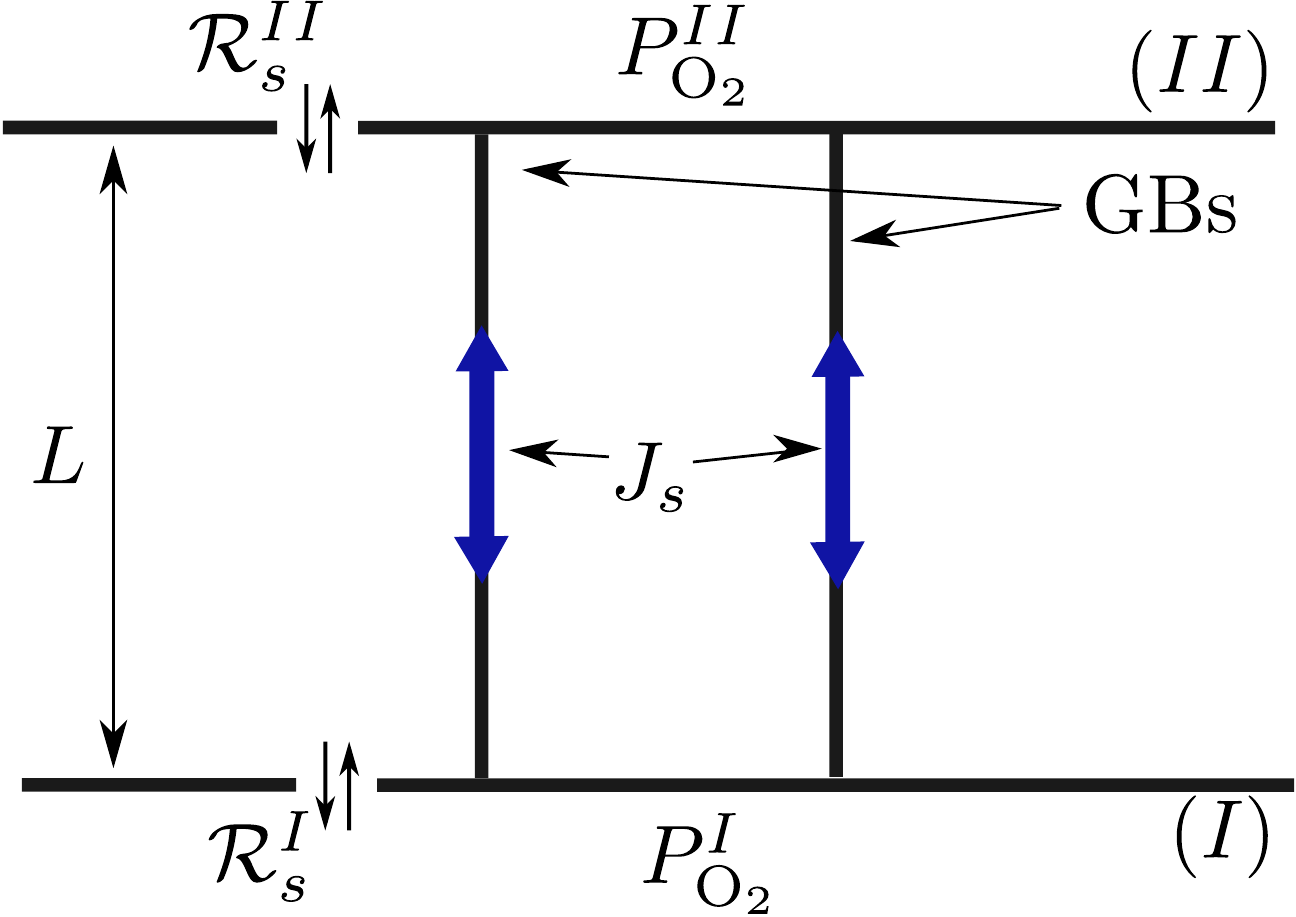} 
     \caption{Schematic of the membrane permeation experiments where mass transfer occurs by grain boundary transport. 
			 Different oxygen gas partial pressures, $\PI$ and $\PII$, are applied on the surfaces. 
			 }
     \label{fig:schematic_slab_gb}
   \end{center}
\end{figure}

Permeation rates of oxygen through a polycrystalline membrane of alumina 
have been reported in the literature~\cite{Kitaoka2009, Kitaoka2016}, and cover a range of oxygen partial pressures.
Scanning Electron Microscope (SEM) imaging of the films prepared under different applied pressures strongly suggests 
that mass transfer occurs along grain boundaries.
The thermodynamic driving force in these experiments is the 
difference in the oxygen chemical potential between the two membrane surfaces
$\Delta \mu_{\text{O}_2} = \kBT\ln\left(\PII/\PI\right)$. 
Figure~\ref{fig:schematic_slab_gb} shows a schematic of membrane permeation experiments.

The experiments included  nominally pure $\alpha$-alumina polycrystals~\cite{Kitaoka2009, Wada2011}, doped
$\alpha$-alumina polycrystals~\cite{Matsudaira2010, Matsudaira2011, Matsudaira2013, Kitaoka2014}, and 
nominally pure $\alpha$-alumina bicrystals~\cite{Matsudaira2011a}, with temperatures of $\sim1700-2000\,\text{K}$.
A simple analysis of the permeation rate data in reference~\cite{Kitaoka2009}, 
assumed a model of one-dimensional, steady-state diffusion, in which either Al or O is transported by vacancy migration, 
depending on the absolute magnitude of the applied oxygen pressure.\\
\indent In the non-doped polycrystalline alumina experiments \cite{Kitaoka2009}, when applying high oxygen pressures 
at surface ($II$), $\Ph = 10^3-10^5\,\text{Pa}$,
while keeping surface ($I$) at $\PI=1\,\text{Pa}$, grain boundary ridges formed on 
the $\Ph$ surface and grain boundary trenches were observed on surface ($I$). 
Applying a low oxygen pressure at surface ($II$), $\Pl=10^{-5}-10^{-8}\,\text{Pa}$, 
while keeping surface ($I$) at $\PI=1\,\text{Pa}$, 
no grain boundary ridges are formed and only grain boundary trenches are observed.
Since the oxygen permeation rates of a single-crystal alumina wafer were below the measurable limit 
and as the visible surface growth and ``dissolution'' proceeds at grain boundaries 
it is reasonable to assume that grain boundaries dominate the transport~\cite{Kitaoka2009}.
The oxygen permeation rates, $P$, for fixed $\PI=1\,\text{Pa}$ 
were found to follow distinct power laws~\cite{Kitaoka2009}; 
in the limit of $\PII=\Ph$,
\begin{align} \label{eqn:powerh}
	P \propto \left(\PII\right)^{3/16},
\end{align}
and in the limit of $\PII=\Pl$, 
\begin{align} \label{eqn:powerl}
	P \propto \left(\PII\right)^{-1/6}.
\end{align}
The power laws and the pressure dependent
formation of the grain boundary ridges have led to the interpretation of the experiments
in terms of, aluminum vacancy transport being dominant in oxygen chemical potential gradients with 
high oxygen pressure magnitude, $\PII=\Ph$,
and oxygen vacancy transport being dominant in the case of $\PII=\Pl$.\\
\indent Indeed, the rational power laws appear in the theory as a direct 
consequence of the +3 and -2 ionic charges of the ions, 
assuming that the negative of these charges is carried by each vacancy, 
with a counter-current of electrons or holes, 
and no time-dependence of the fluxes (the steady-state assumption) or net local charge densities
within the grain boundaries. These assumptions are discussed further in section~\ref{sec:powlaw}.
Furthermore, only aluminum vacancy transport can lead to ridge formation, which is observed by SEM imaging
in the case of $\PII=\Ph$, supporting the above interpretation. 
The switch-over in dominant point defect species in the grain boundary
has been termed ``$p-n$ transition'' in the literature~\cite{Heuer2013,Heuer2011}.\\
\indent The diffusion coefficients determined from alumina bicrystal experiments with 
$\PII/\PI = 10^5\,\text{Pa}/1\,\text{Pa}$ for several distinct grain boundary types 
have been found compatible with those measured in polycrystalline samples~\cite{Matsudaira2011a}.
The bicrystal diffusion coefficients were calculated from the grain boundary ridge volume and a caveat 
regarding this approach is that the formation of the ridges on the $\Ph$ side does not 
necessarily imply an exactly equivalent  mass transport
from the opposite side of the membrane, since oxide can be displaced by formation of internal pores in 
the subsurface region of the crystals, 
as has indeed been observed in some polycrystal permeation experiments~\cite{Matsudaira2010}.


\subsection{Scope of the paper} \label{sec:quest}
The above level of analysis leaves several open questions, 
e.g. does the grain boundary diffusion mechanism, with the associated inhomogeneity of fluxes and electric fields,  
map accurately onto a 1D diffusion problem? 
And what are the magnitude and roles of the surface and interface charges, 
the electric fields,  currents, space charges, 
and transients that are all believed to be present in a three-dimensional film, traversed by grain boundaries?\\
\indent To address these questions we have modelled the transport of oxygen through a planar film 
by making an idealised representation of  the grain structure, in which we suppose the grains to be columnar, 
with identical and perfectly hexagonal cross sections, see figure~\ref{fig:hex_slab}. 
We describe in this paper the set of coupled reaction-diffusion equations we have used to model the 
oxygen permeability across the membrane, our method of solving them, 
and the results we have obtained for the model alumina membrane.
Within this geometry,  
grain boundary transport through the film of charged cation and anion vacancies, electrons, and holes is simulated,
while reactions with the environment constitute the boundary conditions.
We expect our 3D model to be applicable to different materials and to be
readily adaptable to the problem of oxide scale growth,
which is very similar to the problem of oxygen diffusion through a membrane, 
the most significant difference being the boundary conditions.\\
\indent The  equations describing  time-dependent diffusion of ions, electrons and holes through a 
polycrystalline film, driven by electric fields and defect concentration gradients, 
cannot be solved analytically in general, and numerical methods must be applied.
Models including numerical computations to describe transport through films have been
developed for homogeneous films~\cite{Brumleve1978, Fromhold1987, Battaglia1995}, 
for which a one dimensional model may be a suitable approximation.
The symmetry of the present model of idealized columnar hexagonal grains is used to reduce the problem to
2D  boundary diffusion along the rectangular boundaries of the hexagonal grains, 
while explicitly taking account  of the long-ranged electrostatic interactions 
between the charged species within the 3D structure.
In order to be able to describe transient behaviour and  time-dependent environments, 
time-dependent boundary conditions are taken into account,  but the movement of the boundaries is neglected.
This means surface charges can build up or be depleted as a function of time, due to the reactions with oxygen in the environment, in our case a prescribed oxygen partial pressure, 
and the delivery of charged species to the surfaces from the grain boundaries. \\
\indent The system of equations used in our model to describe the fluxes of point defects, 
electrons and holes and their Coulomb interaction, 
is mathematically equivalent to the drift-diffusion (DD) equations applied in 
semiconductor device simulations for electron and hole transport~\cite{Markowich1990, Selberherr1984},
and to the Poisson-Nernst-Planck (PNP) system, which is used for 
ion channel simulations~\cite{Eisenberg1996,Eisenberg1999} 
and other electrochemical applications~\cite{Bazant2004e}.
Unlike most computational  methods of solution for the DD and PNP equations, which solve the Poisson equation in 
differential form and often only consider the steady-state solution,
the solution method developed here allows for time-dependent calculations and 
the Poisson equation is solved in its integral form, taking into account the long-range Coulomb interaction
within the 3D structure.

\section{The hexagonal cell model}

\subsection{2D-periodic hexagonal prism grain structure}

\begin{figure}[H]
\begin{center}
 \includegraphics[width=8cm]{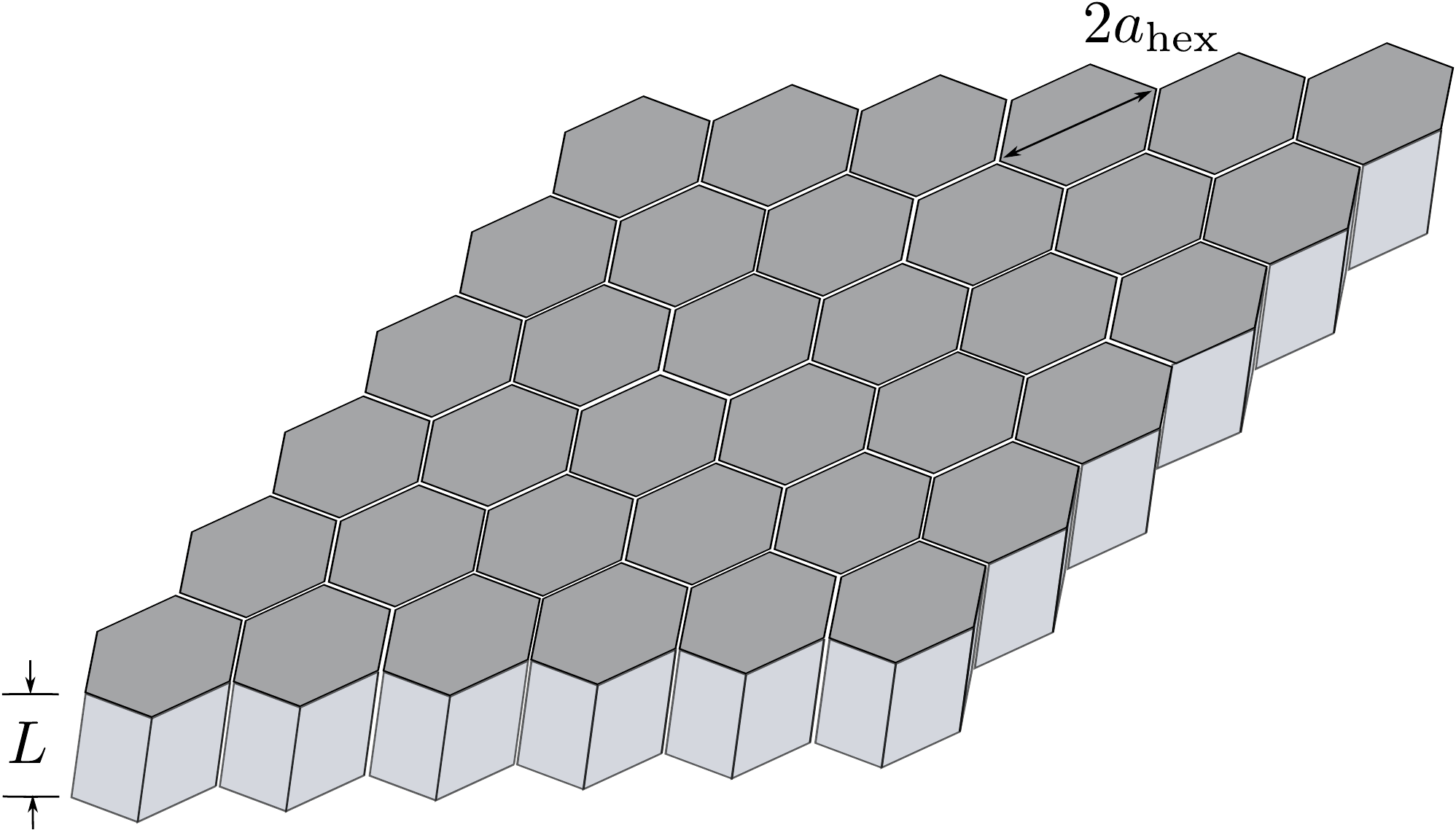}
 \caption{Section of the slab with thickness, $L$, composed of hexagonal cells of side length, $\ahex$.}
	 \label{fig:hex_slab}
\end{center}
\end{figure}

The model is developed for films with columnar grain structure.
The columnar grains are idealized as hexagonal prisms, 
with the rectangular faces denoting the grain boundaries, see figure~\ref{fig:hex_slab}. 
The hexagonal prisms are periodically repeated to construct a slab of infinite extent in two dimensions.\\

Transport through the slab is presumed to be dominated by that through 
the rectangular grain boundaries, located at the interfaces between the cells.
The grain boundaries are assumed to be composed of a very thin homogeneous and isotropic medium of finite width $\delta$. 
The width of the boundary is not a physical width, but rather a theoretical construct, 
which allows concentrations to be expressed per unit  of volume or per atomic site rather than per unit of area.\\
\indent 
From the symmetry of the system only 
the ``irreducible zone'' of the hexagonal cell, 
shown in figure~\ref{fig:hex_prism_irr}, needs to be considered as a domain for calculation.
It includes a triangular piece of the surface hexagon and half of a grain boundary rectangle.
The  2D-periodic tiling enables an explicit calculation of the long-range Coulomb interaction
between the charged point defects.

\subsection{Equations for grain-boundary transport} \label{sec:gb-trans}
The point defect concentrations within the boundary are assumed to be continuous functions of space and time.
Based on the local equilibrium hypothesis the electrochemical potential, $\eta_s$, 
of point defects, and electrons or holes of species ``$s$'', is given by
\begin{align} \label{eqn:loceq}
\eta_s(r,t) = \mu_s(r,t) + z_s e_0 \,\phi(r,t)
\end{align}
where, $\mu_s$ is the chemical potential of species $s$, $\phi$ is the electrostatic potential, 
$z_s$ is the charge (integer number), $e_0$ is the positive elementary charge,
and $r$ is the position vector.
Phenomenologically, the particle flux due to a gradient in the electrochemical potential may be written as,
\begin{align} \label{eqn:flux}
	J_{s} = -\frac{D_s c_s}{\kBT} \nabla \eta_s
\end{align}
which in the ideal solution approximation~\cite{Callen1985} is equivalent to 
\begin{align} \label{eqn:flux}
	J_{s} = -D_s \nabla c_s - \frac{D_s c_s z_s e_0}{\kBT} \nabla \phi
\end{align}
where $D_s$ is the diffusion coefficient, and $c_s(r,t)$ is the concentration (number per unit volume) of species $s$.
Equation~\ref{eqn:flux} is referred to as the Nernst-Planck flux~\cite{Nernst1888,Planck1890}, 
and combines Ohm's law of conduction and Fick's law of diffusion.
It is used here as the constitutive equation for the description of the point defect transport;
magnetic field effects are not considered.\\
\indent The local charge density is given by
\begin{align} \label{eqn:rho}
	\rho(r,t) = \sum_s z_s e_0 \, c_s(r,t)
\end{align}
where the sum is performed over all charged species present.
The instantaneous electrostatic potential can be calculated from the charge density by solving Poisson's equation,
which is given here in integral form
for a linear dielectric material,
\begin{align} 
	\phi(r,t) &= \frac{1}{4 \pi \varepsilon_0 \varepsilon_r} 
	\int_{\mathbb{V}} \frac{ \rho(r',t) }{|r - r'|} d^3r' \label{eqn:phi}
\end{align}
where $\varepsilon_0$ and $\varepsilon_r$ are the vacuum and relative permittivity respectively, 
and the domain of integration $\mathbb{V}$ 
of the ``Coulomb integral'' includes the entire system, in which charge densities are non-zero within a 
slab of infinite extent in two dimensions, composed of identical hexagonal cells and their surfaces.
Overall charge neutrality holds for the domain $\mathbb{V}$, at any instant of time
\begin{align}
	\int_{\mathbb{V}} \rho(r,t) d^3 r = 0.
\end{align}
\indent The continuity equations for the individual species are given by
\begin{align} \label{eqn:conti}
	\frac{\partial }{\partial t} c_s = -\nabla \cdot J_s + R_s
\end{align}
where a reaction term, $R_s$, 
has been added to enable processes such as electron and hole recombination within the grain boundary to be 
described.  
The continuity equations for the different mobile species are used in the simulations to evolve the 
defect concentrations in time; they are solved self-consistently with the 
equation for the electrostatic potential \ref{eqn:phi}, 
which depends on the charge density $\rho(r,t)$. \\
\indent To model coupling effects in the species transport we could 
formulate the dynamics in terms of 
linear irreversible thermodynamics~\cite{Onsager1931a, Pottier2009}, but this theory does not provide explicit
expressions for the constitutive equations including the transport coefficients.
The linear constitutive equations, like Fick's first law, have a phenomenological basis; 
however, they can also be thought of as laws of inference based on probability theory~\cite{Grandy2008}.
For electron and hole transport in semiconductor device simulations a system of equations, 
mathematically equivalent to equations ~\ref{eqn:flux}, \ref{eqn:phi}, \ref{eqn:conti} 
was first introduced by van Roosbroeck~\cite{VanRoosbroeck1950}. They are referred to as drift-diffusion equations, 
and can be derived from the Boltzmann transport equation 
by either the Hilbert expansion or the moment method~\cite{Markowich1990,Selberherr1984}.

\subsection{Time dependent boundary conditions} 

\begin{figure}[H]
	\centering
	\begin{minipage}[b]{0.4\linewidth}
		\includegraphics[width=\textwidth]
		{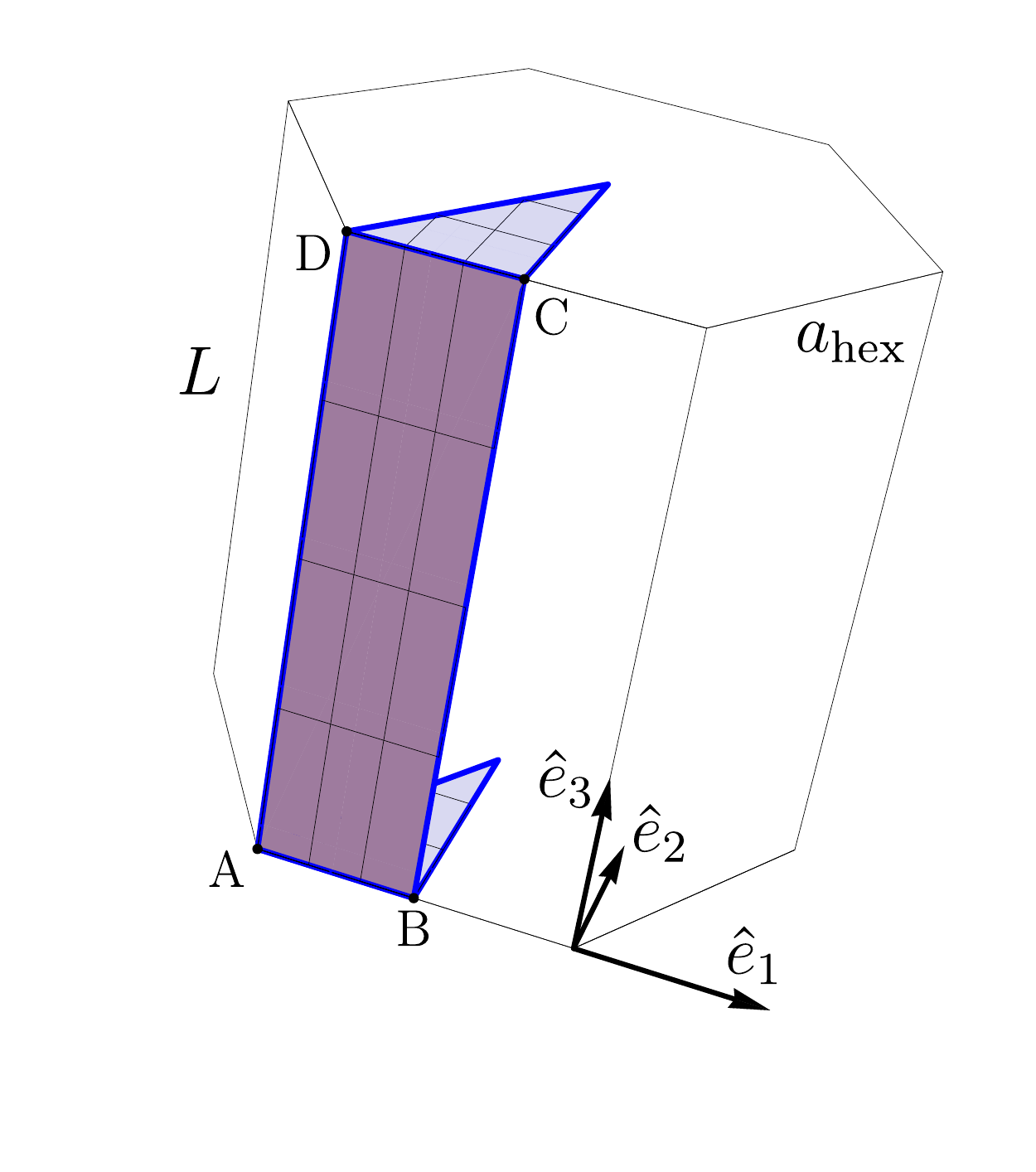}
		 \caption{Irreducible zone of the hexagonal cell. 
		 The grain boundary rectangle, enclosed by $\overline{ABCD}$, is indicated. The coordinate system is defined.}
     \label{fig:hex_prism_irr}
 \end{minipage}
 \quad
 \begin{minipage}[b]{0.35\linewidth} 
     \includegraphics[width=\textwidth]
		 {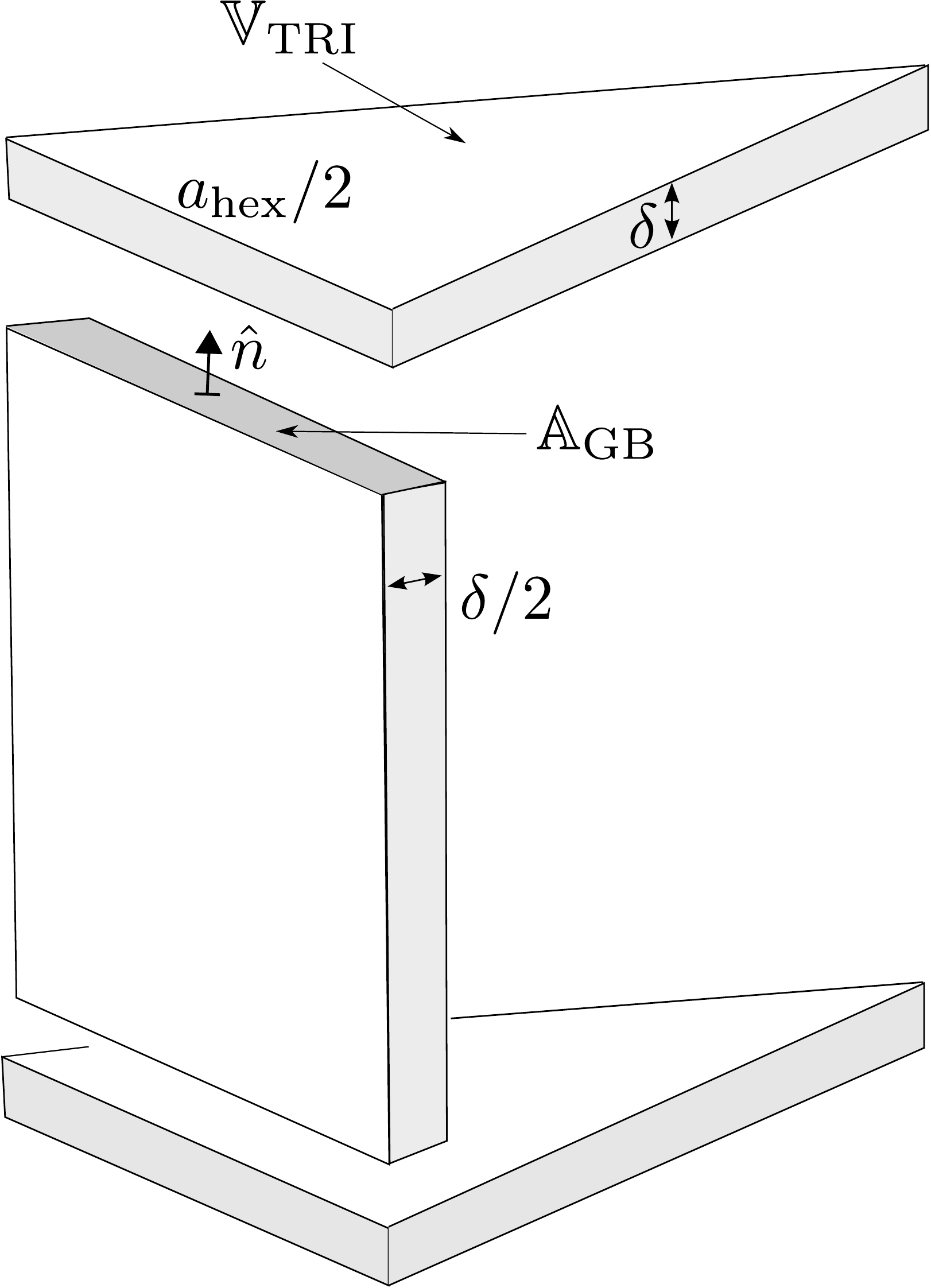}
	\caption{Exploded-view of the irreducible zone of the hexagonal lattice
			 defining the surface and grain boundary domains. 
			 The grain boundary thickness, $\delta$, which is equal to the surface layer thickness, $\delta$, 
			 the grain boundary cross-section, $\AGB$, the outward pointing unit normal, $\hat n$, 
			 and the volume of the surface triangle, $\mathbb{V}_{\text{TRI}}$, are indicated. 
		 Drawn out of proportion, since in general $\delta \ll \ahex$. }
     \label{fig:wedge_expl}
	\end{minipage}
\end{figure}

The boundary conditions are formulated to describe oxide creation and dissolution at the slab surfaces by
reaction with the environment.
We refer to our system, which includes its upper and lower surfaces, as `the slab'. 
The thin surface layers of the slab
are treated as a homogeneous and isotropic medium of  thickness $\delta$.
The flux of defects between the 
 surfaces and the grain boundary, and the reactions between the slab and the 
 environment,  change the  concentrations of species $s$ in the surface layers, $\mathcal{C}_s$, as expressed by
\begin{align} \label{eqn:BC}
	\int_{\mathbb{V}_{\text{TRI}}} \frac{\partial \mathcal{C}_s}{\partial t} \;d^3 r 
	= \int_{\AGB} J_s \cdot \hat{n} \;d^2 r + \int_{\mathbb{V}_{\text{TRI}}} \mathcal{R}_s  \;d^3 r .
\end{align}  
Reactions  between the slab surface and the environment produce species $s$ in the surface layer at a rate $\mathcal{R}_s$.
The unit normal, $\hat{n}$, and the integration domains are defined in figure~\ref{fig:wedge_expl}. 
Equation \ref{eqn:BC} holds for all species ``s'' and separately for 
the surfaces ($I$) at $x_3=0$ and ($II$) at $x_3=L$.

\indent Assuming transport to be much faster across the surface than between surface and grain boundary, 
uniform defect concentrations are used on the surfaces. 
This leads to the following simplification of equation \ref{eqn:BC}
\begin{align} \label{eqn:BCsimp}
	\frac{\partial }{\partial t} \mathcal{C}_s
	=  \frac{1}{\mathbb{V}_{\text{TRI}}}\int_{\mathbb{A}_{GB}} J_s \cdot \hat{n} \;d^2 r + \mathcal{R}_s
\end{align}
which is used as the boundary condition in the simulations reported below. 
If the surface transport mechanism and parameters are known it is straightforward to relax 
the above assumption.\\
\indent No separate boundary condition is needed for the electrostatic potential since it is calculated by 
a summation technique, only requiring the instantaneous 
charge distribution as a function of position.
The reaction rates in the grain boundary, $R_s$, and on the surface, $\mathcal{R}_s$, 
depend on the application of the model.
For the present  purpose of describing the alumina permeation experiments the rate equations are 
derived with the law of mass action, 
and discussed in more detail in section~\ref{sec:reac}. 
The application to oxide scale growth will require a separate boundary condition at the interface between oxide and metal. \\
\indent The initial conditions also depend on the application. For the oxygen permeation experiments 
they are discussed in section~\ref{sec:init}.

\section{Method of solution}

An object oriented C++ code has been developed to solve the
system of coupled partial differential equations~\ref{eqn:flux},~\ref{eqn:conti}, and~\ref{eqn:BCsimp},
self-consistently 
with the Coulomb integral, equations~\ref{eqn:rho}, and~\ref{eqn:phi}, which is approximated by the 
summation technique described in section~\ref{sec:sum}.
Since the system of equations is nonlinear, involves vastly different rates in the diffusion processes and the
reactions, and involves different length scales characterized by the Debye length, defined below, and the system size, 
this is a challenging computational problem.

\subsection{Dimensionless equations}\label{sec:scal}

To obtain dimensionless equations for the numerical calculations 
the variables, parameters, and fields are scaled as follows~\cite{Selberherr1984}:
\begin{equation} \label{eqn:scaling}
	\begin{aligned}
	r &= \frac{1}{L} \; \widetilde{r}, \hspace{0.5cm}
	D_s = \frac{1}{\Dref} \; \widetilde{D}_s, \hspace{0.5cm}
	c_s(r,t) = \frac{1}{\cref} \; \widetilde{c}_s, \\ 
	t &=  \frac{\Dref}{L^2} \; \widetilde{t}, \hspace{0.5cm}
	\phi(r,t) = \frac{e_0}{\kBT} \; \widetilde{\phi},   
	\end{aligned}
\end{equation}
where L is the thickness of the scale; $\Dref = \max\{D_s\}$ is the largest diffusion coefficient for all species;
$\cref$ is a suitable reference concentration; 
and $e_0$ is the elementary positive charge. 
To avoid unnecessarily heavy notation, the original quantities were denoted with the ``tilde'' mark over the symbol
in the definition of the scaling, and the dimensionless quantities without the tilde are used in the following. \\
\indent With the scaling defined by equations~\ref{eqn:scaling}, 
the system of transport equations and the Coulomb integral can be brought into dimensionless form
\begin{subequations}\label{eqn:sdimless}
\begin{align} 
	\frac{\partial }{\partial t} c_s &= -\nabla \cdot \left( -D_s \nabla c_s - D_s z_s c_s \nabla \phi \right) \\
	\kappa^{2} \phi(r,t) &= 
	\frac{1}{4 \pi} \int_{\mathbb{V}} \frac{ \sum_s z_s c_s(r',t) }{|r - r'|} d^3r' \label{eqn:scaledphi} 
\end{align}
\end{subequations}
where the dimensionless parameter, $\kappa$,
\begin{align}
	\kappa = \frac{\lD}{L}, \hspace{1cm} 
	\lD = \left( \frac{\varepsilon_0 \varepsilon_r \kBT}{e_0^2 \, \cref} \right)^{1/2} \label{eqn:lD}
\end{align}
and the reference screening length, $\lD$, are introduced. 
$L$ denotes the thickness of the slab, see figure~\ref{fig:hex_slab}.
The Debye screening length is defined by
\begin{align} \label{eqn:Debye}
	\LD = \left( \frac{\varepsilon_0 \varepsilon_r \kBT}{e_0^2 \sum_s z_s^2 c_s^0}\right)^{1/2}
\end{align}
where $c_s^0$ are the spatially uniform concentrations obtained in the limit of $T\rightarrow\infty$,
while holding the total number of each species, $N_s$, constant.
In the application considered in this work the oxygen gas chemical potentials are fixed at the surfaces,
the concentrations, $c_s(r,t)$, are independent variables, 
and the total number of each species depends on time, $N_s(t)$; 
however, $\cref$ is chosen such that $\lD$ has similar magnitude to $\LD$, 
and $\lD$ is therefore referred to as the reference screening length.

Overall charge conservation and zero total charge within  the system are maintained during the 
evolution of the concentrations, 
while charge is redistributed within the grain boundary and moved in or out of the surfaces by the fluxes.

\subsection{Discretization of the transport equation}

The finite difference method is used to discretize 
the continuum equations in space and time~\cite{Selberherr1984,LeVeque2007}.
For the spatial discretization a rectangular mesh is used, and it covers the 
irreducible zone of the hexagonal cell, indicated in figure~\ref{fig:hex_prism_irr}. 
The concentrations and electrostatic potential 
at mesh node $(i,j)$ at position $r(x_1^i,x_3^j)$ in the grain boundary ($x_2=0$) 
and discrete time $t_n$ are denoted by, $c_s(x_1^i,x_3^j,t_n)=c_s^{i,j,n}$,
and $\phi(x_1^i,x_3^j,t_n) = \phi^{i,j,n}$, respectively, where $i = 1,2,...,N_1$ and $j=1,2,...,N_3$.\\
\indent The mesh spacings near the surfaces need 
to be significantly smaller than the reference screening length to resolve the behaviour near the surfaces correctly.
However, the whole grain boundary cannot be meshed with such a fine spacing in the $\hat{e}_3$ direction,
because the summation technique used 
for calculating the Coulomb interaction would become too computationally expensive.
Therefore, a layer-adapted mesh is used in the $\hat{e}_3$ direction.\\ 
\indent The Nernst-Planck flux in equation~\ref{eqn:flux} is 
approximated with the Scharfetter-Gummel discretization 
scheme~\cite{Scharfetter1969};
in one dimension for fixed $j$ it is given by
\begin{align}
	&J_s^{i+1/2,j,n} = \nonumber\\
	&-D_s
	\frac{c_s^{i+1,j,n} \; \text{B}\left(-z_s\Delta \phi^{i,j,n} \right) - c_s^{i,j,n} \; 
\text{B} \left(z_s \Delta \phi^{i,j,n} \right) }{\Delta x_1^i}
\end{align}
where $J_s^{i+1/2,j,n}$ is the flux of species $s$  between node $i$ and $i+1$ at time $t_n$, 
$\Delta x_1^i= x_1^{i+1}-x_1^i$ is the length of the interval, 
$\Delta \phi^{i,j,n}=\phi^{i+1,j,n}-\phi^{i,j,n}$ is the potential difference between the mesh nodes at time $t_n$, and 
$\text{B}(x)=x/(\exp(x)-1)$ is the Bernoulli function~\cite{Selberherr1984}.\\
\indent For the time stepping the continuity equation~\ref{eqn:conti} is discretized in implicit form, 
\begin{align}
	&c_s^{i,j,n}-c_s^{i,j,n-1} + \frac{\Delta t_n}{\Delta \overline{x}^i_1 } \, 
	\left(J_s^{i+1/2,j,n} - J_s^{i-1/2,j,n}\right) + \\
	&\frac{\Delta t_n}{\Delta \overline{x}^j_3 } \, 
	\left(J_s^{i,j+1/2,n} - J_s^{i,j-1/2,n}\right)  - \Delta t_n R_s^{i,j,n}= 0 \nonumber
\end{align}
where $\Delta \overline{x}^i_1 = (\Delta x^i_1+\Delta x^{i-1}_1)/2$, 
$\Delta \overline{x}^j_3 = (\Delta x^j_3+\Delta x^{j-1}_3)/2$, and $\Delta t_n = t_{n}-t_{n-1}$.
The initial time step, $\Delta t_1$, is chosen such that the discretizations in space and time 
have a similar order of accuracy, hence
	$\Delta t_1 \approx (\Delta x_{\mathrm{min}})^2$
where $\Delta x_{\mathrm{min}}$ denotes the smallest mesh element. 
$\Delta x_{\mathrm{min}}$ is chosen to have the same size as the surface layer thickness, $\delta$,
therefore typically $\Delta t_1 = 10^{-6}$ and the time step is subsequently increased adaptively, to improve 
efficiency of the procedure while ensuring convergence at each time step. The maximum time step 
is typically $\Delta t_{\mathrm{max}}\sim 0.1$ which allows calculations to be performed long enough to reach 
steady-state conditions without the need for exceptional computational resources.

\subsection{Calculation of the long-range Coulomb interaction} \label{sec:sum} 

\begin{figure}[H]
	\centering
	\begin{minipage}[b]{0.35\linewidth}
     \includegraphics[width=\linewidth]{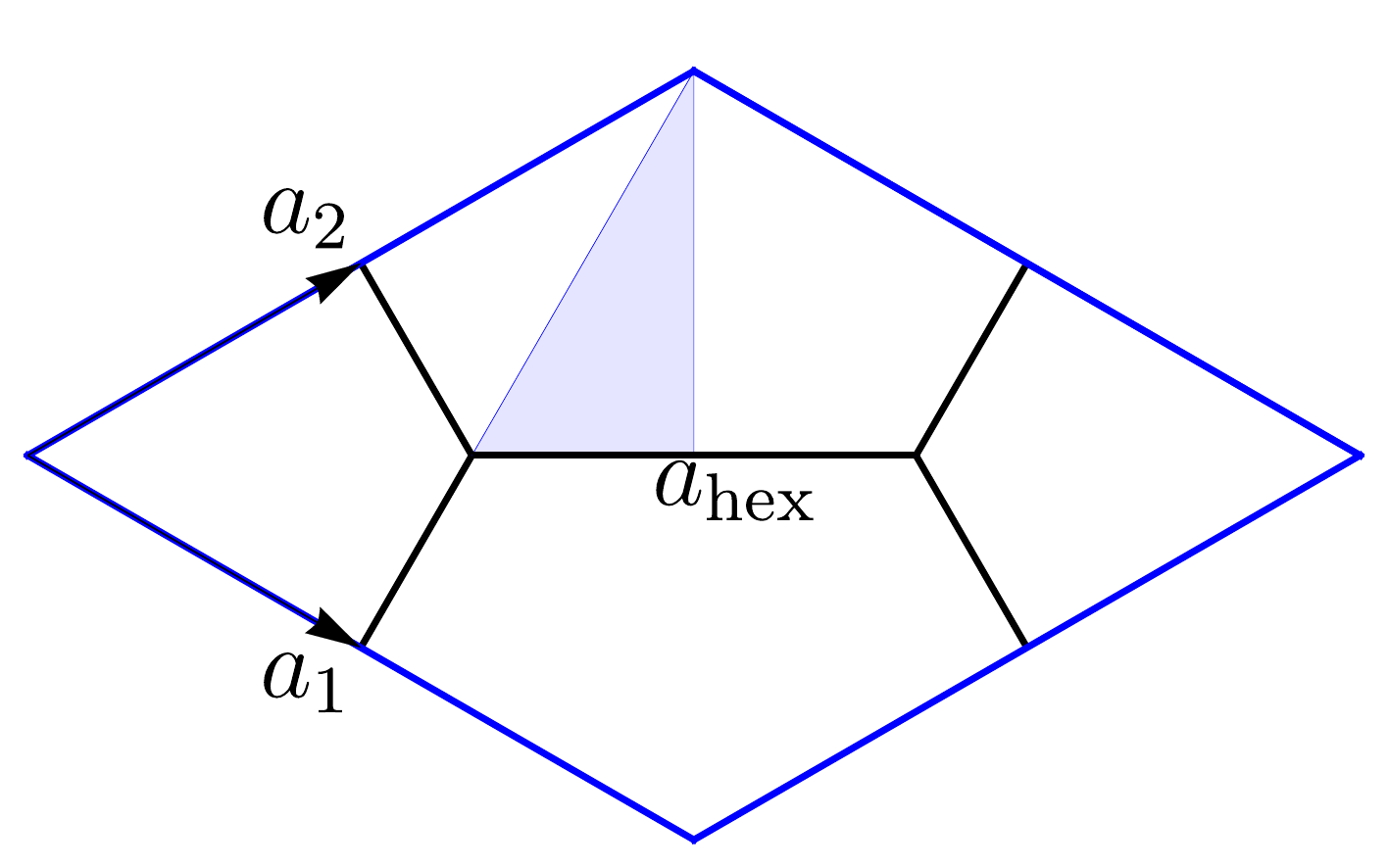}
     \caption{A 2D repeat unit of the hexagonal tiling used in the simulations. The shaded 
		 area is the irreducible zone of the surface, referred to as the surface triangle.}
     \label{fig:2d_repeat_unit}
\end{minipage}
\quad
\begin{minipage}[b]{0.35\linewidth}
     \includegraphics[width=0.85\linewidth]{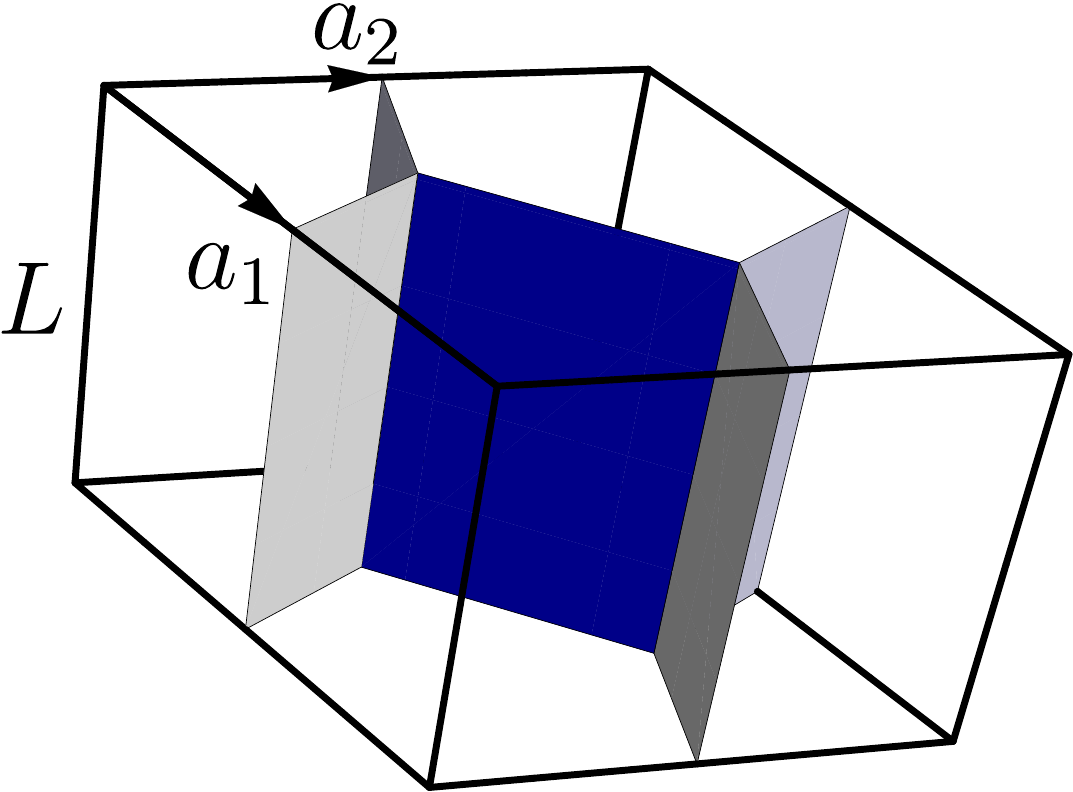}
		 \centering
     \caption{A 3D repeat unit of the infinite slab, seen in plan view in figure~\ref{fig:2d_repeat_unit}.
		 The shaded planes indicate the grain boundaries within the repeat unit.}
     \label{fig:hexagonal_system}
  \end{minipage}
\end{figure}

Since the functions $c_s(r,t)$ are discretized in space and time 
the Coulomb integral in equation~\ref{eqn:scaledphi} can be approximated with a summation technique.
In this section the concentration at the mesh nodes is abbreviated as 
$c_s(r_j,t_n) = c_s(x_1^{j_1},x_2^{j_2},x_3^{j_3},t_n)=c_s^{j,n}$, with the 
composite index $j=(j_1,j_2,j_3)$, and in the same way $c_s(r_i,t_n) = c_s^{i,n}$.
The volume element corresponding to mesh node $j$ is denoted by, $\nu_{j}$. The charge density 
in the volume, $\nu_j$, around mesh node $j$ is turned into a point charge, 
$q^{j,n} = \sum_s z_s \, c_s^{j,n} \, \nu_j$, placed at position $r_j$.
With this definition the Coulomb integral is converted into the Coulomb sum
\begin{align}
	\phi(r,t_n) = \kappa^{-2} \; \sideset{}{'}\sum_j^{\mathbb{V}} \frac{\sum_s z_s \, c_s^{j,n} \, \nu_{j}}{|r - r_j|}
\end{align}
where $\phi(r,t_n)$ is still a continuous function of space, 
and it can only be evaluated at the 
discrete times, $t_n$, since the $c_s^{j,n}$ are only known at discrete times $t_n$.
The prime indicates that the possible term $r=r_j$ is excluded from the summation.
The summation index $j$ runs over all volume elements of the infinite slab, and since the potential decays with $r^{-1}$, 
the sum is only slowly and conditionally convergent; it cannot be trivially truncated.
Therefore, the Parry summation technique~\cite{Parry1975,Parry1976}, 
which is an Ewald summation technique for 2D periodic systems, is used.
In this  technique the sum is split into a real and a Fourier (reciprocal) space part, 
the short-range interactions 
are evaluated in real space and the long-range interactions are evaluated in Fourier space. 
The advantage is rapid convergence compared to the direct summation.\\
\indent The rhombus shown in figure~\ref{fig:2d_repeat_unit} is defined as the repeat unit of the 2D periodic tiling, 
and the rhombohedral prism shown in figure~\ref{fig:hexagonal_system} 
is a repeat unit of the infinite slab, used to carry out the Parry summation.
The symmetry of the hexagonal prism, with distinct hexagonal surfaces on the top and bottom surface,
is used to increase the computational efficiency of the evaluation of the Coulomb sum.

\indent During a calculation  the charge density changes with time and 
the potential needs to be updated with the changing charge density;
however, the geometry of the hexagonal structure and the mesh do not change. 
Therefore, the Parry summation is performed only once, at the beginning of the calculations, 
to determine the Green's function for the given mesh and periodic cell structure. 
The Green's function is calculated for unit charges on the individual mesh nodes, 
and is stored as a matrix, $g_{ij}$, for the nodes of the irreducible zone, 
where $g_{ij}$ denotes the potential at node $i$ due to a unit charge and all its images, generated by symmetry and 
periodicity, at node $j$.
During the time-dependent calculations, 
the potential is calculated by summing the discretized charge density, multiplied by the Green's function, 
over the nodes of the irreducible zone, 
\begin{align}
	\phi(r_i,t_n) = \sum_j q^{j,n} g_{ij}\,.
\end{align}
This strategy greatly reduces the computational cost of evaluating the Coulomb integral.\\
\indent A stretched mesh with variable mesh spacings, which are different in the $\hat e_1$ and $\hat e_3$ direction,
is found to lead to divergence problems when performing the summation over the point charges, $q^{j,n}$.
Therefore, ``Gaussian smearing'' is used for the charge densities at the individual mesh nodes. 
The local charge density at mesh node $j$ and discrete time $t_n$, $\rho(r_j,t_n)$, 
is replaced by a normalized Gaussian distribution
\begin{align}
	\rho_g(r,r_j,t_n) = \rho(r_j,t_n) \, \left(\frac{\beta}{\pi}\right)^{\frac{3}{2}} \, \exp(-\beta |r-r_j|^2) 
	\hspace{0.4cm} 
\end{align}
where $\beta = (2\sigma^2)^{-1}$ and $\sigma$ is the width of the Gaussian centered around $r_j$.
With the Gaussian distribution the electrostatic potential contributed
by the charge density corresponding to mesh node $j$ takes the form
\begin{align}
	\phi_g(r,r_j,t_n) =  \kappa^{-2} \frac{ q^{j,n} }{|r - r_j|} \text{erf}(\sqrt{\beta} |r - r_j|)  .
\end{align}
In the limit of $|r-r_j| \rightarrow 0$ the electrostatic potential contributed
by the charge density centred on mesh node $j$ itself becomes
\begin{align}
	\phi_g(|r-r_j| \rightarrow 0, t_n) =  2 \,  \kappa^{-2}   \left( \frac{\beta}{\pi} \right)^{ \frac{1}{2} } q^{j,n} ,
\end{align}
which has to be added to the potential at this node generated by all the other charges in the system. 
The width $\sigma$ of the Gaussians is chosen to be that of the smallest mesh spacing in the slab.

\subsection{Iterative solver for the self-consistent calculations}

A multidimensional Newton method is used to solve the nonlinear system of discretized equations, 
including the continuity equations for all species on all mesh points, their boundary conditions,
and the Coulomb summation to obtain the self-consistent electrostatic potential value at each mesh point.
Depending on the number of mesh points and the number of species the system,
which needs to be solved in every time step,
can contain several thousand variables. 
Therefore, we have implemented a Jacobian-Free-Newton-Krylov (JFNK) method
from the \textsf{NOX} package of the Trilinos Project~\cite{Heroux2005} in our code.
JFNK methods are nested iteration methods and 
can achieve Newton-like convergence without the cost of forming and 
storing the true Jacobian required for ordinary Newton methods~\cite{Knoll2004}.

\section{Diffusion of oxygen through an alumina film}\label{sec:diff-ox}
The effective charges of the point defects are defined with respect to the ions of the pristine lattice, 
and the Kr\"oger-Vink notation for point defects is used, 
although for generality we use integers to denote the charge states rather than the 
original superfix ${}^\bullet$  or ${} '$ symbols, but retain the ${}^\times$ notation for neutral species.

Oxygen membrane permeation or diffusion through an oxide film is a non-equilibrium process, involving 
oxygen exchange and electron transfer reactions at the oxide surfaces 
and the transport of defects between the surfaces.

\subsection{1D analytic model for the permeation rate} \label{sec:powlaw}

%

Our derivation of
an analytic model for the oxygen permeation rate emphasizes the role of the electric field 
for the establishment of approximately 
stoichiometric proportions for the dominant vacancy and electronic species of the surface concentrations, $\mathcal{C}_s$,
which determine the permeation rate,
and the appearance of an effective (ambipolar) diffusion coefficient, see also~\ref{app:mod-deriv}, 
in the presence of the self-consistent electric field.

The concentrations of the defect species on the surfaces exposed to the oxygen gas environment,
characterized by the oxygen partial pressure and the temperature,
evolve over time and depend on the reactions taking place and the transport to and from the surfaces.
At the oxide -- gas interface the absorption (desorption) of oxygen 
and creation (annihilation) of aluminum vacancies can be described by
\begin{align} \label{eqn:RM}
	\frac{1}{2} \text{O}_2(g) + 2 \eprime 
	&\underset{k_{\text{Al},b}}{\stackrel{k_{\text{Al},f}}{\rightleftharpoons}}  \OO + \frac{2}{3} \VAl  . 
\end{align}
For the purpose of a simple analytical treatment we assume
a steady-state is reached at the surfaces, in which
Schottky and electron-hole equilibrium, see reactions~\ref{eqn:RS} and~\ref{eqn:Rel}, 
characterized by the equilibrium constants $K_{\text{S}}$ and $K_{\text{eh}}$, respectively, are attained.
The reaction~\ref{eqn:RM} could be formulated with holes instead of electrons, however, 
assuming instantaneous equilibration between electrons and holes both formulations yield identical results.
At high applied oxygen gas partial pressures, aluminum vacancies are formed predominantly at the surface and 
oxygen vacancies are annihilated, with the reverse scenario applying for low oxygen partial pressures.
Since instantaneous Schottky equilibration is assumed at the surfaces no additional reaction is required involving
oxygen incorporation by oxygen vacancy annihilation.
In reaction~\ref{eqn:RM} for $\Ph$ applied 
electrons are consumed by the oxygen atoms creating oxygen sublattice sites and aluminum vacancies,
and due to the electron hole equilibration the concentration of holes increases simultaneously; therefore,
in the limit of $\Ph$ applied, aluminum vacancies and holes are the dominant species on the $\Ph$ surface.
Similarly, oxygen vacancies and electrons are expected to dominate on a $\Pl$ surface.
The electric field due to the charged defect species modifies the transport to and from a particular surface 
in such way that local charge neutrality holds 
approximately between the dominant vacancy species, $\nu$, 
and the charge compensating electronic species, $\nuo$, on the surface
\begin{align}
	|z_{\nu}| \mathcal{C}_{\nu} \simeq \mathcal{C}_{\nuo}
\end{align}
where the $\{\nu,\nuo\}$ pair denotes $\{\VAl,\hdot\}$ at high oxygen partial pressure,
and $\{\VO,\eprime\}$ at low oxygen partial pressure. 
This approximation does not hold in general for intermediate pressures,
or if the vacancies are much more numerous than the electronic defects in equilibrium.
Applying the law of mass action to the reaction given in~\ref{eqn:RM} with equilibrium constant $K_{\text{Al}}$,
together with Schottky and electron hole equilibrium at the surfaces, 
the concentration of the dominant vacancy species, $\nu$, is found to adhere to the simple power law given by
\begin{align}
	&\mathcal{C}_{\nu}  = f_{\nu} \, \POtwo^{\; n_{\nu}} \label{eqn:cpower}\\
	&n_{\nu} = \frac{-z_{\nu}}{2|\zO| \left(|z_{\nu}|+1\right)} \label{eqn:nnu}\\
	&f_{\nu} = \left(\left|{z_{\nu}}\right|^{-|z_{\nu}|} K_{\text{S}}^{3-|z_{\nu}|} K_{\text{eh}}^{|z_{\nu}|(|z_{\nu}|-2)} K_{\text{Al}}^{-z_{\nu}/2} \right)^{\frac{1}{|z_{\nu}|+1}}
\end{align}
where $\zO$ is the integer charge number of the oxygen ion, 
$z_{\nu}$ is the charge of the dominant vacancy species, 
and the power law exponent, $n_{\nu}$, which in our case takes the values, $n_{\VO}=-1/6$ 
and $n_{\VAl} = 3/16$, and the prefactor, $f_{\nu}$, are introduced.  
It is noteworthy that  the realisation of these power laws in the experiments supports 
the model of vacancies carrying nominal ionic charges, 
not the fractional charges that are usually estimated in electronic structure calculations, 
which typically vary from 1.0 to 1.6 in DFT calculations for oxides~\cite{Dovesi1992, Batyrev1999}. 
It is also far from obvious that simple point defect diffusion, well understood in bulk crystals,  
is the mechanism of diffusion in grain boundaries, 
in which the prefactors of diffusion coefficients are anomalously large~\cite{Heuer2008,Harding2003}.

 
In general, the flux of aluminum ions, $\JAl$ implies the take up or release of $| 3\JAl/ 4|$ molecules of $\text{O}_2$
per unit time and unit area of surface, and
the opposite flux of oxygen ions leads to $|\JO/2|$ molecules $\text{O}_2$ taken up or released.
The ionic charges have the opposite sign and 
the fluxes have the opposite sign due to their different gradients, 
therefore the permeation rate in molecules of oxygen per unit area per second is given by
\begin{align}
	P &= \left| \frac{3}{4}\JAl \right| + 
	\left| \frac{1}{2}\JO \right| =  \frac{1}{4 e_0} |\Iion|, \nonumber
\end{align}
and hence by using $\Iion = \Ivac$
\begin{align}
	P &= \frac{1}{4 e_0} |\Ivac|\label{eqn:perm-Ivan}
\end{align}
where the current density carried by the ions, $\Iion$, and vacancies, $\Ivac$,
represent the same physical current density, which has to be balanced by the electron and hole currents.\\
\indent To interpret and compare with results of our fully time-dependent approach, we introduce a simple one-dimensional,
steady-state treatment, the derivation of which is given in~\ref{app:mod-deriv}.
By considering only the flux of the dominant vacancy species, denoted by suffix $\nu$, and its charge-compensating
electronic species, $\nuo$:
\begin{align}
	\Ivac \simeq I_{\nu},
\end{align}
the permeation rate is approximated by
\begin{subequations}
\label{eqn:analytic-model}
	\begin{align} 
		P &= \frac{|z_{\nu}|}{4} D_{\nu}^{\text{eff}} \frac{|\mathcal{C}_{\nu}^{II}-\mathcal{C}_{\nu}^{I}|}{L}\label{eqn:perm},\\
	D_{\nu}^{\text{eff}} &= \frac{D_{\nu} D_{\overline{\nu}}}{D_{\nu}|z_{\nu}|+D_{\overline{\nu}}}  (|z_{\nu}|+1)\label{eqn:Deff}.
\end{align}
\end{subequations}
Using equation~\ref{eqn:cpower} 
the permeation rate can be written as
\begin{align} \label{eqn:perm-rate}
	P &= \frac{|z_{\nu}|}{4} D_{\nu}^{\text{eff}} f_{\nu} \, 
	\frac{|\left( \PII \right)^{n_{\nu}} - \left( \PI\right)^{n_{\nu}}|}{L}
\end{align}
where again $n_{\VO}=3/16$, and $n_{\VAl}=-1/6$.
Equation~\ref{eqn:perm-rate} shows that the permeation rate becomes a power law only 
for $\left(\PII\right)^{n_{\nu}}\gg \left(\PI\right)^{n_{\nu}}$. It is also worth noting that the permeation rate
is not proportional to the oxygen pressure difference across the membrane, 
but rather to the difference in the vacancy concentration between the two surfaces.\\
\indent The oxidation rate ``$J$'' of aluminum in atoms per second per unit area in terms of the permeation rate
of oxygen in molecules per second per unit area is given by
\begin{align} \label{eqn:ox-perm}
	J = \frac{2}{3} P  \,.
\end{align}
\indent We can compare the above results with the Wagner theory~\cite{Wagner1933, Atkinson1985, Hauffe1965},
in which the oxidation rate is given by
\begin{align} \label{eqn:wagner}
	J = \frac{1}{|\zAl |\zO^2 e_0^2} \int_{I}^{II}
	\frac{(\sigma_{\VO}+ \sigma_{\VAl})(\sigma_h+\sigma_e)}{\sigma_{\VO} + \sigma_{\VAl}+\sigma_h + \sigma_e} d\muO
\end{align}
where $\zAl$ is the aluminum ion charge and $\muO$ is the oxygen component chemical potential.
By assuming that the conductivity of the dominant vacancy species, $\sigma_{\nu}=D_{\nu} c_{\nu} z_{\nu}^2 e_0^2/\kBT$,
is much larger than that of the other vacancy species, and similarly for the dominant electronic species, $\nuo$,
the oxidation rate simplifies to
\begin{align}
	J = \frac{1}{|\zAl| \zO^2 e_0^2} \int_{I}^{II}
	\frac{ \sigma_{\nu} \sigma_{\overline{\nu}} }{ \sigma_{\nu} + \sigma_{\overline{\nu}} } d\muO \,.
\end{align}
Applying the charge neutrality approximation to this equation and considering relation~\ref{eqn:ox-perm} 
the permeation rate given in equations~\ref{eqn:analytic-model} follows,
which makes our 1D model consistent with the Wagner theory.
The derivation of this form of the Wagner theory, however, requires Schottky and electron-hole equilibrium
at any position in the film or scale,
\begin{align} \label{eqn:eq}
	3\mu_{\VO} + 2 \mu_{\VAl} = 0 \;\; \text{and} \;\; \mu_{\eprime}+\mu_{\hdot} = 0
\end{align}
neither of which is enforced in our treatment. 
Strictly speaking, the above conditions only need to be met by the spatial variations of the chemical 
potentials, but if equations~\ref{eqn:eq} hold at the surfaces both formulations are equivalent
and there is no arbitrary additive constant.

\indent Equations~\ref{eqn:analytic-model} are equivalent to the formulas given by~\cite{Kitaoka2009} for the permeation rate
in the limit of $D_{\nu}\ll D_{\overline{\nu}}$.
However, the details of the derivation are different 
such as the assumption of purely 
conductive transport of electrons and holes, $I_s= - \sigma_s \nabla \phi$, which is made in~\cite{Kitaoka2009}.

\subsection{Time-dependent 3D calculations} \label{sec:PL}

In the calculations within the hexagonal slab model 
transport of vacancies is simulated for the 3D grain structure and grain boundaries
with finite width $\delta$.
The time-dependent calculations prove to have a similar 
steady-state limit to the 1D model described above, 
in which the divergences of the fluxes and current densities vanish. 
The vacancy flux is calculated directly from the concentration and electrostatic potential 
gradients, and the permeation rate follows from equation~\ref{eqn:perm-Ivan}.
Before the steady-state is achieved, an average permeation rate can be calculated 
by numerical integration of the vacancy flux through the grain boundary, which converges faster 
to the steady-state permeation rate than the permeation rate calculated from the vacancy fluxes on individual 
mesh nodes.

\subsubsection{Reaction equations} \label{sec:reac}

Exchange of oxygen between the gas phase and the oxide surface includes multiple steps, namely
adsorption, dissociation, surface diffusion, charge transfer, and incorporation into the oxide surface,
each of which might be the rate limiting step.
In the simulations the assumptions of instantaneous equilibration between the vacancy species inducing the Schottky 
equilibrium, and the instantaneous equilibration between electrons and holes are relaxed, 
and the respective processes are formulated with rate dependent reaction equations.
In addition to the oxygen incorporation mechanism given in~\ref{eqn:RM},
we consider oxygen absorption (desorption) by annihilation (creation) of an oxygen vacancy
\begin{align} \label{eqn:RO}
	\frac{1}{2} \text{O}_2(g) + \VO &\underset{k_{\text{O},b}}{\stackrel{k_{\text{O},f}}{\rightleftharpoons}} \OO + 2 \hdot . %
\end{align}
The Schottky reaction is included at the alumina - oxygen gas surfaces
\begin{align} 
	\text{Nil}& \underset{k_{\text{S},b}}{\stackrel{k_{\text{S},f}}{\rightleftharpoons}} \frac{2}{3} \VAl + \VO. \label{eqn:RS} 
\end{align}
Electron and hole recombination and generation,
\begin{align} \label{eqn:Rel}
	\text{Nil}\underset{k_{\text{eh},b}}{\stackrel{k_{\text{eh},f}}{\rightleftharpoons}} \eprime + \hdot, 
\end{align}
is considered at the grain boundary and on the surfaces.\\
\indent In this application the surface reactions, $\mathcal{R}_s$, 
depend on the time-dependent surface concentrations, $\mathcal{C}^{I/II}_s(t)$, and 
the oxygen partial pressures, $\PI$ or $\PII$, 
on the respective surfaces, ($I$) or ($II$),
and the reaction rate constants, $k_{i,f}$ and $k_{i,b}$
\begin{align}
	&\mathcal{R}_s^{\text{j}} \left( \,
		\{ \mathcal{C}^{\text{j}}_s(t)\}; \POtwo^{\text{j}}, \{ k_{i,f}\}, \{k_{i,b} \} \, \right)\\
		&s \in \{ \VO, \VAl, \eprime, \hdot \} \; \text{species} \nonumber\\
		&i \in \{\text{O},\text{Al},\text{S},\text{eh}\} \hspace{0.5cm} \text{see reactions~\ref{eqn:RO},
	\ref{eqn:RM}, \ref{eqn:RS}, \text{and} \ref{eqn:Rel}} \nonumber\\
	&\text{j} \in \{ I, II \} \; \text{surface index} \nonumber
\end{align}
The law of mass action is used to derive expressions for the reactions $\mathcal{R}_s$, and $R_s$.
The coupled system of reactions, $\mathcal{R}_s$, at the alumina - oxygen gas surfaces, is used for the
boundary conditions of the transport equations. The equations are provided in~\ref{app:reac} 
and they describe reactions that may proceed in either direction, depending on the 
species concentrations and the oxygen gas pressure applied.

\subsection{Simulation results}

Results are presented here to address some of 
the questions posed in section~\ref{sec:quest} of the introduction, regarding the membrane experiments~\cite{Kitaoka2009} 
and their interpretation with 1D diffusion models based on the Wagner theory.
Time dependent calculations are performed and characteristic aspects of the dynamics are highlighted 
and explained. A permeation calculation reproduces the power laws found in~\cite{Kitaoka2009}.

\subsubsection{Initial conditions and choice of parameters} \label{sec:init}

The system at $t=t_0$ is assumed to be of strictly stoichiometric composition, $2c_{\VO}(t_0) = 3c_{\VAl}(t_0)$, 
and the set of initial values of the concentrations, $\{c_s(t_0)\}$, and $\{\mathcal{C}_s(t_0)\}$,
is assumed to be in equilibrium with oxygen partial pressure $\Peq$, 
which is defined as the reference pressure so that only ratios $\POtwo/\Peq$ enter the equations.
The initial concentrations in the surface layer and in the grain boundary are chosen 
to be equal, $c_s=\mathcal{C}_s$, and independent of the position in the grain boundary and on the surface. 
This choice also constrains the electron and hole concentrations, $c_{\eprime}(t_0)=c_{\hdot}(t_0)$, 
in order for overall charge neutrality to be maintained.
The initial values for the concentrations, $c_s(t_0)$, are used to specify 
the equilibrium constants, $K_i$, of the reactions, which are also related to the reactions rates
\begin{align} \label{eqn:law-MA}
	K_i &= \exp\left( \frac{\sum_s w_{i,s} \mu_s^0}{\kBT} \right) = \prod_s c_s^{w_{i,s}}(t_0) = \frac{k_{i,f}}{k_{i,b}} 
\end{align}
where $\mu_s^0 = -\kBT \,\text{ln}(c_s(t_0))$ is the reference chemical potential
with $c_s(t_0)$ in units of $\cref$, and
$w_{i,s}$ is the stoichiometric coefficient of 
species $s$ in the $i$-th reaction.

This choice of initial parameters only leaves undetermined
the ratio of point defect to electronic defect initial concentrations, 
which is a function of the difference between 
the vacancy formation (segregation) energy and the Fermi level.
For bulk $\alpha$-alumina the ionic and electronic disorder has been analysed from experimental thermodynamic data
\cite{Mohapatra1978} and more recently from first principles calculations~\cite{Ogawa2014}. 
At temperature $1900\,$K electronic disorder is found dominant for the bulk material. 
However, we are unaware of data for the surface and interface equilibrium defect concentrations
and aim to justify our choice by comparison to the experimental permeation data, see section~\ref{sec:perm}.
We expect the point defects which are favourably formed at the surfaces and interfaces
to introduce states in the band gap, and therefore assume much higher vacancy, electron, and hole
concentrations at the surfaces, and interfaces than in the bulk material.
The reference concentration for the simulations is chosen as $\cref = 10^{18}\,\text{cm}^{-3}$,
which would correspond to about $4\times10^{-5}$ defects per formula unit of bulk $\alumina$,
and will only be reached in the high or low pressure limits.
The concentrations are initialized at $t=t_0$ with $c_{\hdot} = 0.1\, \cref$, and $c_{\VO} = 5\times10^{-3} \, \cref$.

For $T=1900\,$K and with $\varepsilon_r^{\alumina} = 9.8$, the reference screening length, 
see equation~\ref{eqn:lD}, becomes, $\lD = 9.4 \,$nm.
The thickness of the slab is set to, $L = 1\, \mu$m, the side length of the hexagon to 
$\ahex = 1 \, \mu$m, the grain boundary and surface layer thickness to $\delta=1\,$nm.
The diffusion coefficients of the species are chosen as: 
$D_{\hdot}=D_{\eprime}=1 \, \Dref$, and $D_{\VO}= D_{\VAl}= 0.01\, \Dref$ where $\Dref$ 
is unknown and used to define the time scale of the simulations. 
One reaction rate constant in each reaction has to be estimated and they 
are set to: $k_{\text{O},b} = 10^3$, $k_{\text{Al},b} = 10^2$,
and $k_{\text{eh},b}= k_{\text{S},b} = 10^3$.

\indent In sections~\ref{sec:evol}, \ref{sec:rho-phi} 
and~\ref{sec:charg-dyn} calculations are discussed for which the oxygen partial pressure 
is raised at surface ($II$), the pressure at surface ($I$) is kept constant at $\PI=\Peq$, and
the pressure ratio is given by $\PII / \PI = 10^5$. 
Snapshots of the evolution at two times, $t_n$, and $t_N$, with $t_n\ll t_N$, are shown.
The time $t_n$ is chosen to capture characteristic behaviour in the initial transient,
and $t_N$ is the time at which steady-state conditions are achieved.\\

\subsubsection{Evolution of the concentrations} \label{sec:evol}

\begin{figure}[H]
	\centering
	\includegraphics[width=9cm]{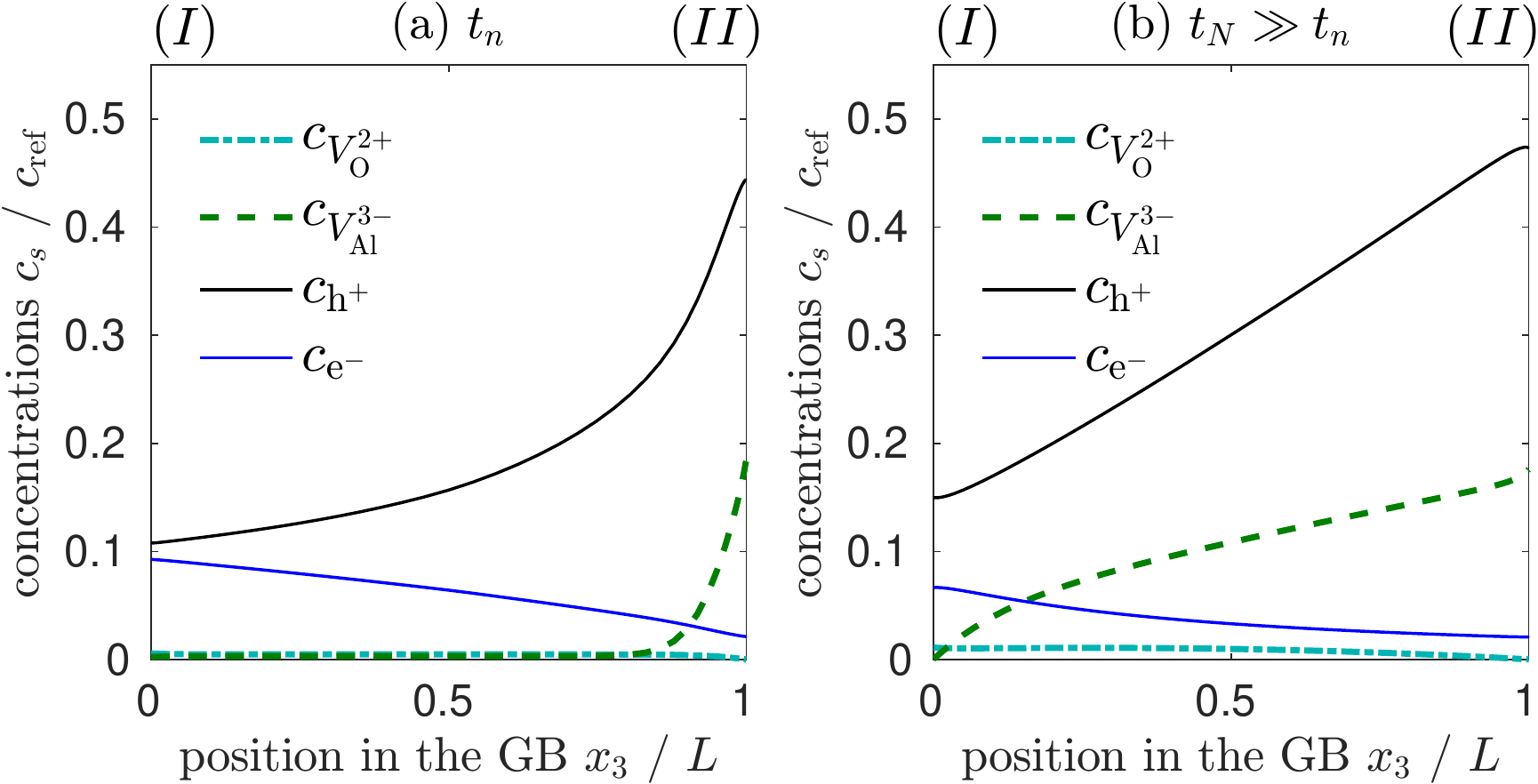}
	\caption{Snapshots of the concentrations of vacancies, electrons, and holes as a function of $x_3$ in the grain boundary plane, 
	with fixed $x_1=\ahex/4$,
	for two different times and with $\PII/\PI=10^{5}$. Scaled units with $\cref=10^{18}\,\text{cm}^{-3}$ and $L=1\,\mu\text{m}$.}
	\label{fig:1-conc}
\end{figure}

Figure~\ref{fig:1-conc} shows snapshots of the concentrations;
the coordinate system is defined in figure~\ref{fig:hex_prism_irr}.
The pressure $\PII=\Ph$ and the $\VAl$ and the $\hdot$ concentrations 
are the dominant species. Local charge neutrality holds approximately for the 
dominant species, $3c_{\VAl} \approx c_{\hdot}$, throughout most of the grain boundary and at surface ($II$) 
in the steady-state $t_N$ limit but is clearly violated at surface ($I$).

\subsubsection{Evolution of the charge density and the electrostatic potential} \label{sec:rho-phi}

\begin{figure}[H]
	\centering
	\includegraphics[width=10cm]{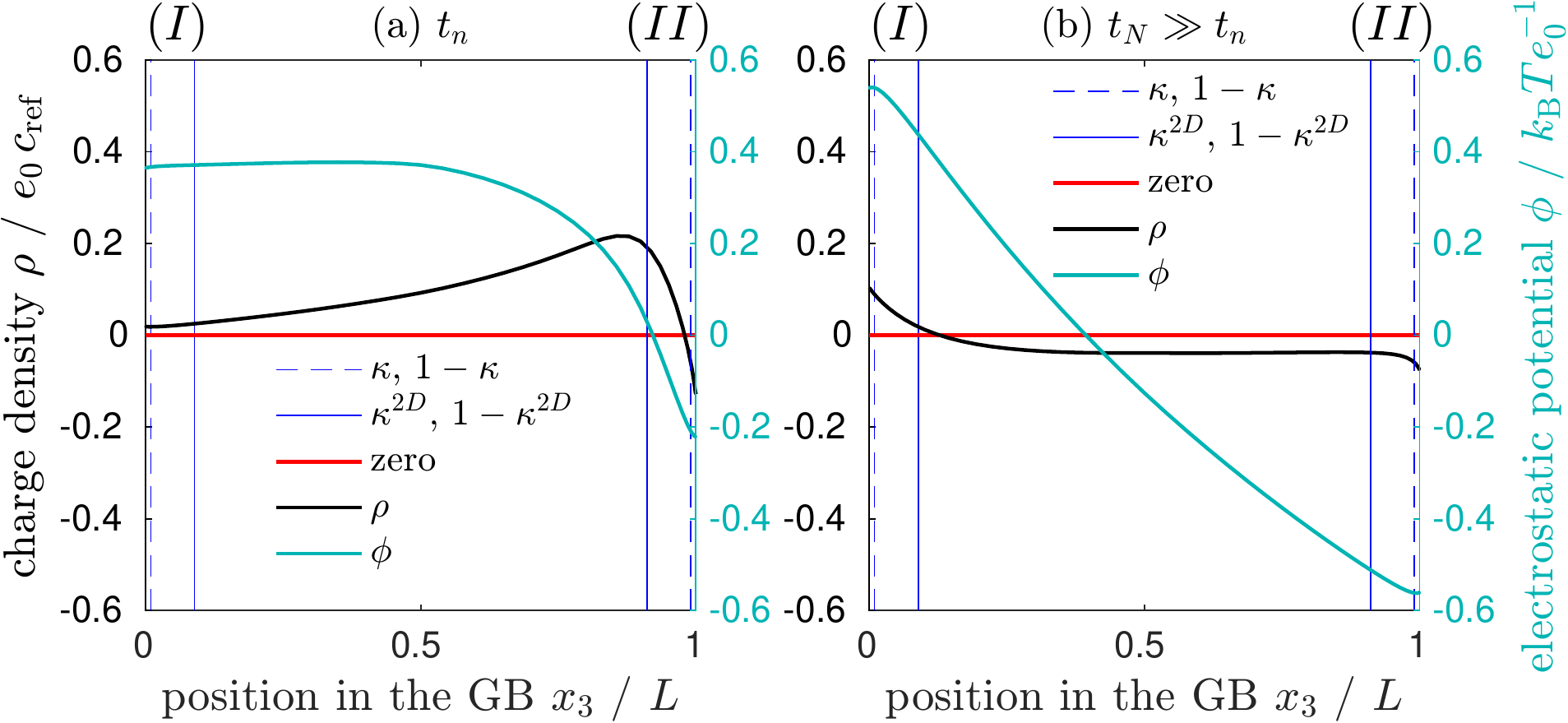}
	\caption{Snapshots of the charge density and the electrostatic potential as a function of $x_3$ at fixed $x_1 = \ahex/4$
		for two different times. 
		Two screening lengths $\kappa$, and $\kappa^{2D}$ are indicated and described in the text.  
 Scaled units with $e_0\,\cref= 0.16\,\text{C}\,\text{cm}^{-3}$, $\kBT/e_0 = 0.16\,\text{V}$, 
 and $L=1\,\mu\text{m}$. The stages (a), and (b) of the overall dynamics are discussed in the text.}
	\label{fig:1-rho-phi_1D}
\end{figure}

Figure~\ref{fig:1-rho-phi_1D} shows the charge density, $\rho$, and the electrostatic potential, $\phi$, 
corresponding to the concentrations shown in figure~\ref{fig:1-conc}.
Charge is accumulated near the high pressure surface ($II$) 
initially until time $t_n$, see figures~\ref{fig:1-rho-phi_1D} (a) 
and~\ref{fig:iz-charg}; this change in the local charge density is due to the
different magnitudes of the diffusion coefficients of the mobile species and the requirement of charge conservation
within the irreducible zone of the structure. 
If all species had the same diffusion coefficient the charge density would be identically zero, $\rho(t) \equiv 0$, 
for all times in calculations with the present model. 
The negatively charged surface ($II$) generates a linearly decreasing potential with increasing 
$x_3$ ($\phi= -|a| x_3+b$);
however, at $t_n$ it is screened to an almost constant value by the charge density that has accumulated 
near surface ($II$). At time $t_N$, see figure~\ref{fig:1-rho-phi_1D} (b), 
some of charge in the grain boundary has propagated to surface ($I$) and the grain boundary has become 
weakly negatively charged, see also figure~\ref{fig:iz-charg}. However, the electrostatic potential has become 
almost linear as a function of $x_3$, and is dominated by the surface charges; the remaining departure from
linearity is due to the charge density within the grain boundary.

The quantity $\kappa$ denotes the scaled reference screening length $\lD/L$, defined in equation~\ref{eqn:lD},
and characterizes the decay in the local 
charge density from the surface into the grain boundary. 
$\kappa^{2D}$ is the equivalent of the scaled reference screening length for two dimensional systems
and is defined here as
\begin{align}\label{eqn:lD2d}
	\kappa^{2D}=\lD^{2D}/L, \;\; \text{where} \;\; \lD^{2D} = \frac{\varepsilon_0 \varepsilon_r \kBT}{e_0^2 \cref \delta}.
\end{align}
The simulations show that $\lD^{2D}$ is a better estimate for the spatial extent of the variations in the 
charge density within the grain boundary than $\lD$. 
It should be pointed out here that simulations 
in which the grain boundaries and surfaces are idealized as planes without finite thickness $\delta$
would yield the same results, 
and $\cref \delta$ in equation~\ref{eqn:lD2d} would be replaced by a reference concentration per unit area.
The only requirement for this to hold
is that the total numbers of each species present in the irreducible zone initially,
are chosen equal for the simulations with and without finite thickness $\delta$.

%

\subsubsection{Surface and grain boundary charges} \label{sec:charg-dyn}

\begin{figure}[H]
 \begin{center}
	 \includegraphics[width=6.2cm]{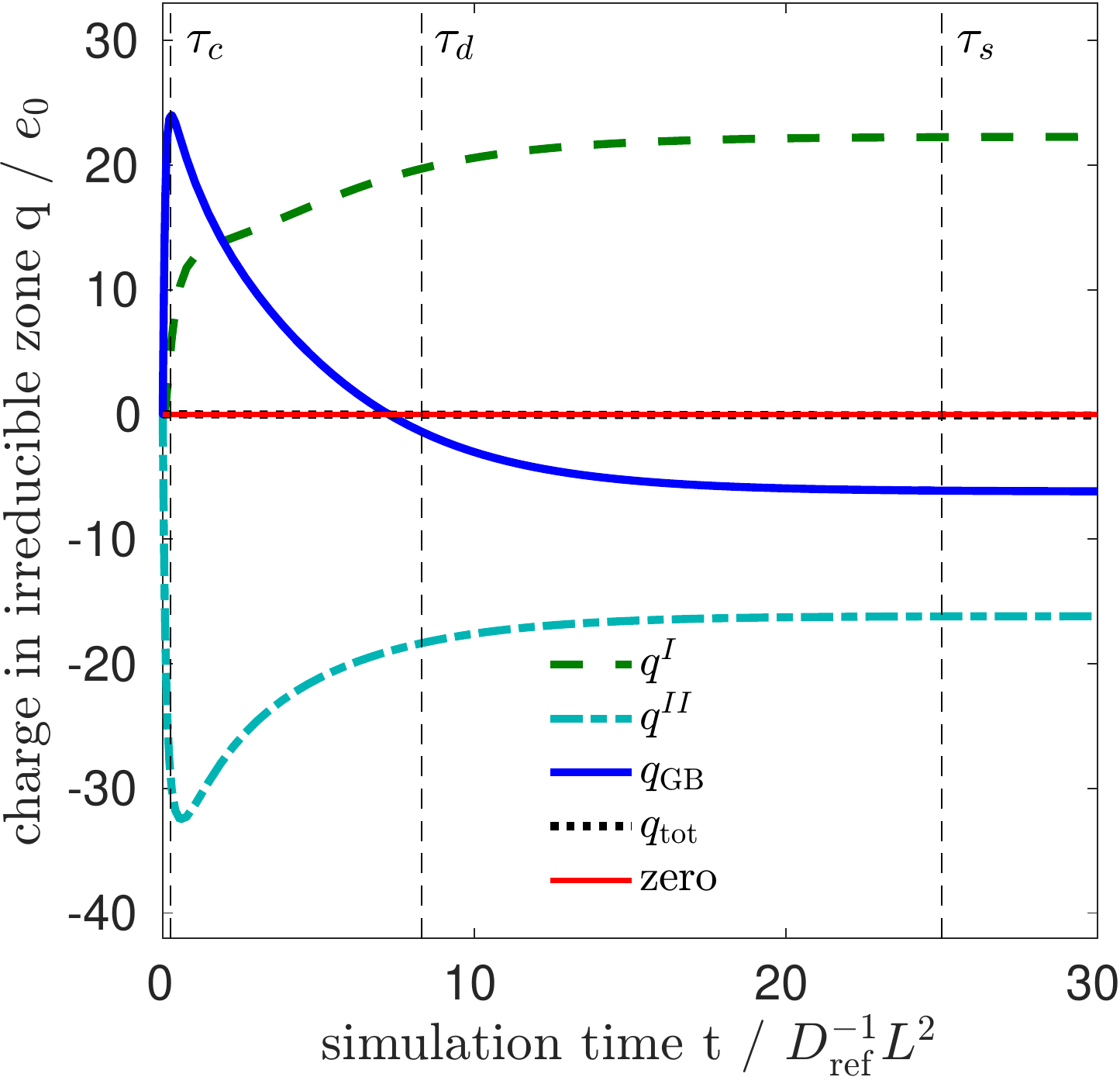} 
  \end{center}
	\caption{The charge (integrated charge density) in the irreducible zone as a function of time.
		$q^I$ and $q^{II}$ are the charges in the surface triangles of surfaces ($I$) and ($II$), respectively,
		$\qgb$ is the charge in the grain boundary, $q_{\text{tot}} =q^I+q^{II}+\qgb$ 
		is the total charge in the irreducible zone.
		The characteristic time scales $\tau_c$, $\tau_d$, and $\tau_s$ are indicated, and further discussed in the text.
Scaled units are used with the thickness of the slab, $L=1\,\mu$m, and $e_0$ the positive elementary charge.}
	\label{fig:iz-charg}
\end{figure}

The simulations are initialized with zero total charge and charge conservation
requires the total charge in the irreducible zone of the structure, $q_{\text{tot}}$, to remain equal to zero.
This is satisfied for the calculations performed, see figure~\ref{fig:iz-charg}, which provides a useful check
on the numerical accuracy and stability of the solution.

The charge integrated over the grain boundary volume changes with time, $\qgb(t)$, 
and the boundary 
carries an excess of electrons in the steady-state limit. 
This demonstrates that local charge neutrality, which is an assumption of the simple 1D models, 
is not consistent with this 3D model.
Two characteristic time scales for the charging and discharging of the 
grain boundary are found. The time constants can be estimated by analogy to ``$RC$''-circuits, with 
time constant $\tau=RC$, where $R$ is the resistance, and $C$ is the capacitance. 
The initial charge built up within the grain boundary is due to the diffusion of the faster of the dominant species, 
for $\PII=\Ph$ the positively charged holes, $\nuo=\hdot$, and the time constant is estimated as follows from appropriate
values of $R$ and $C$:
\begin{align}
	R_{\nuo} &=  \frac{1}{\langle \sigma_{\nuo} \rangle }\frac{4L}{\ahex \delta},\\
	C_{\text{hex}} &= \varepsilon_0 \varepsilon_r \frac{\ahex^2 \sqrt{3}}{8L}, \\
	\tau_c &
	\sim  \frac{\varepsilon_0 \varepsilon_r}{\langle \sigma_{\nuo}\rangle} \frac{\ahex \sqrt{3}}{2 \delta}. \label{eqn:tauc}
\end{align}
The time constant characterizing the discharging of the grain boundary is estimated from the 
diffusion of the slower of the dominant species, for $\PII=\Ph$ the aluminum vacancies, $\nu=\VAl$:
\begin{align}
	R_{\nu} &=  \frac{1}{\langle \sigma_{\nu} \rangle }\frac{4L}{\ahex \delta},\\
	\tau_d &
	\sim  \frac{\varepsilon_0 \varepsilon_r}{\langle \sigma_{\nu}\rangle} \frac{\ahex \sqrt{3}}{2 \delta}. \label{eqn:taud}
\end{align}
Where the angle brackets denote the spatial average and 
the physically meaningful conductivity is the product $\sigma \delta$.
A third time scale characterizes the time elapsed until reaching the steady-state, and is estimated here
from the diffusion length using the effective diffusion coefficient defined in equation~\ref{eqn:Deff}.
For the parameters uses here, $D_{\nu}=D_{\nuo}/100$ and in the 
limit of $D_{\nu} \ll D_{\nuo}$,
\begin{align}
	\tau_s \sim \frac{L^2}{D_{\nu}^{\text{eff}}} \simeq \frac{L^2}{(|z_{\nu}|+1) D_{\nu}} \label{eqn:taus}.
\end{align}
The three time scales are indicated in figure~\ref{fig:iz-charg}.

\subsubsection{Membrane permeation calculations} \label{sec:perm}

\begin{figure}[H]
 \begin{center}
	 \includegraphics[width=7.5cm]{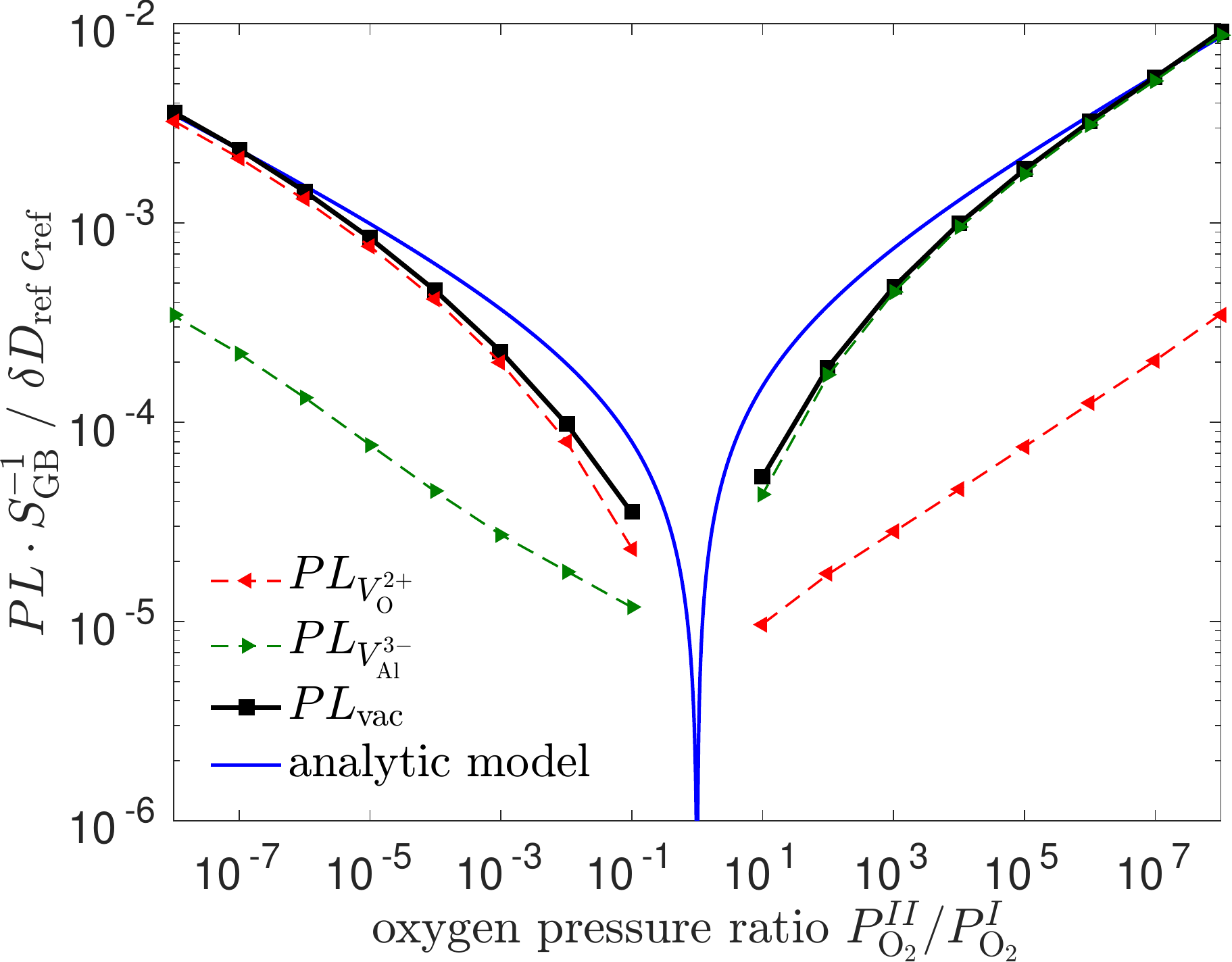}
  \end{center}
	\caption{Oxygen permeation rate in the steady-state limit as a function of the ratio of applied oxygen gas pressures. 
		The simulated permeation rate tends to the same power law exponents 
		as those determined in the 1D analytic model (blue line), $n_{\VAl}=3/16$ and $n_{\VO} = -1/6$, 
		see section~\ref{sec:powlaw}, and the experimentally found ones~\cite{Kitaoka2009}.
		$\delta$ is the grain boundary width, $\Dref\,[\text{m}^{2}\, \text{s}^{-1}]$, $\cref=1.66 \,\text{mol}\,\text{m}^{-3}$,
	and $S_{\text{GB}}=4/\sqrt{3}\ahex$ is the grain boundary density of the hexagonal cell structure.}
	\label{fig:PL}
\end{figure}

The permeation rate normalized by the thickness of the slab, $PL$, 
can be calculated from the current densities of the mobile defect species. 
Figure~\ref{fig:PL} shows the permeation rate in the steady-state limit 
as a function of the oxygen pressure ratio $\PII/\PI$ for a range of $\PII$ values, while $\PI$ is held fixed. 
The individual contributions of the $\VAl$ and $\VO$ are indicated 
by $PL_{\VAl}$ and $PL_{\VO}$, respectively.
The parameters are the same as those in the previous section.

Comparing the simulated permeation rate with the experimental values~\cite{Kitaoka2009, Wada2011} for
$T=1900\,$K and $\PII/\PI=10^5\,\text{Pa}/1\,\text{Pa}$ the reference diffusion coefficient is calculated, 
$\delta \Dref = 3.4\times 10^{-13}\,\text{m}^3\,\text{s}^{-1}$.
From the average aluminum vacancy concentration in the 
grain boundary at $\PII/\PI=10^5$, $\langle c_{\VAl} \rangle = 10^{17}\,\text{cm}^{-3}$, and the concentration of 
alumina formula units, $c_{\alumina}=2.26\times10^{22}\,\text{cm}^{-3}$,
the aluminum diffusion coefficient 
is estimated to be 
$\delta D_{\text{Al}} = \delta D_{\VAl} \,\langle c_{\VAl} \rangle /2c_{\alumina}
= 7.6\times 10^{-21}\,\text{m}^3\,\text{s}^{-1}$, 
which is close to  the value 
reported in~\cite{Kitaoka2009}, $\delta D_{\text{Al}}= 4.5\times 10^{-21}\,\text{m}^3\,\text{s}^{-1}$. 
The agreement indicates that the reference concentration is a reasonable choice provided 
the assumptions of $c_{\nu}(\Peq) \ll c_{\nuo}(\Peq)$ and $D_{\nu} \ll D_{\nuo}$ are applicable.
Indeed, if electrons and holes would diffuse slower than the vacancies, $D_{\nu} \gg D_{\nuo}$, the 
permeation rate would be limited by the electronic defects and the diffusion coefficients determined 
from the permeation experiments~\cite{Kitaoka2009} would reflect the electronic diffusion coefficient
rather than the ionic one.

The variation of the logarithm of the permeation rate $P$, see equation~\ref{eqn:perm-rate}, 
with the oxygen chemical potential at surface ($II$) in
the limits of high and low applied oxygen partial pressure, $\PII$, yield the power law exponent corresponding to the 
dominant defect species, $\nu$,
\begin{align} \label{eqn:P-n}
	\frac{\partial \, \text{ln} \,P }{ \partial \, \text{ln} \,\PII} = n_{\nu}
\end{align}
in the limit of high and low $\PII$.

\begin{figure}[H]
 \begin{center}
	 \includegraphics[width=6.7cm]{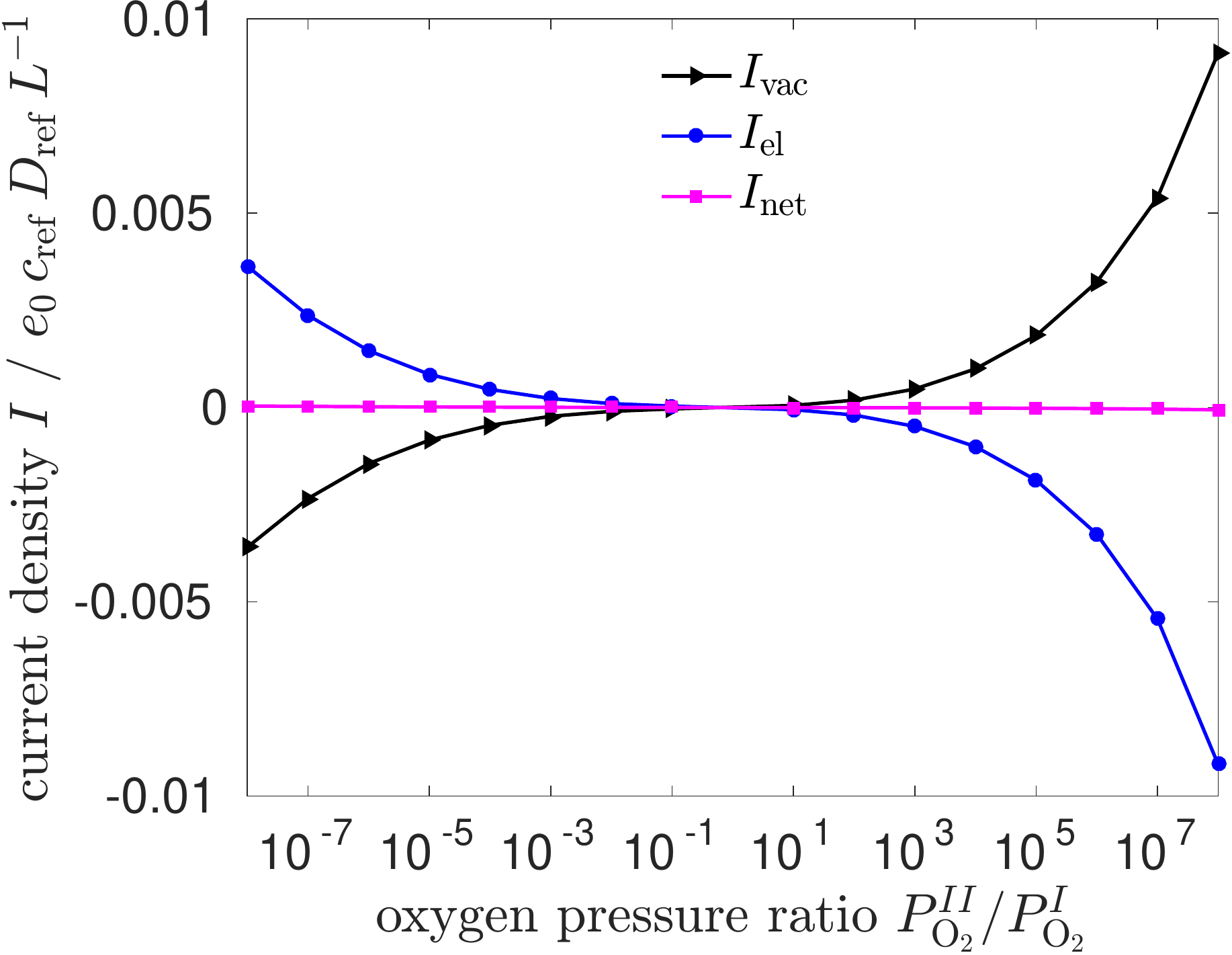}
  \end{center}
	\caption{Current densities per grain boundary ``cross-section area'' $\AGB$, see figure~\ref{fig:wedge_expl},
		in the steady-state limit as a function of the ratio of applied oxygen gas pressures.
	Scaled units with $e_0 \, \cref = 0.16\,\text{C}\,\text{cm}^{-3}$, and $L=1\,\mu\text{m}$.}
	\label{fig:I}
\end{figure}

\indent An asymptotic electronic current density, $\Iel$, and vacancy current density, $\Ivac$
can be calculated similarly to the permeation rate. 
The current density of vacancies, $\Ivac=I_{\VO}+I_{\VAl}$, 
is equal and opposite to the electronic current, $\Iel = I_{\eprime}+I_{\hdot}$, at all pressure
ratios in the long time limit, see figure~\ref{fig:I}, this means the net current, $\Inet=\Ivac+\Iel$ is zero.

\begin{figure}[H]
 \begin{center}
	 \includegraphics[width=7.5cm]{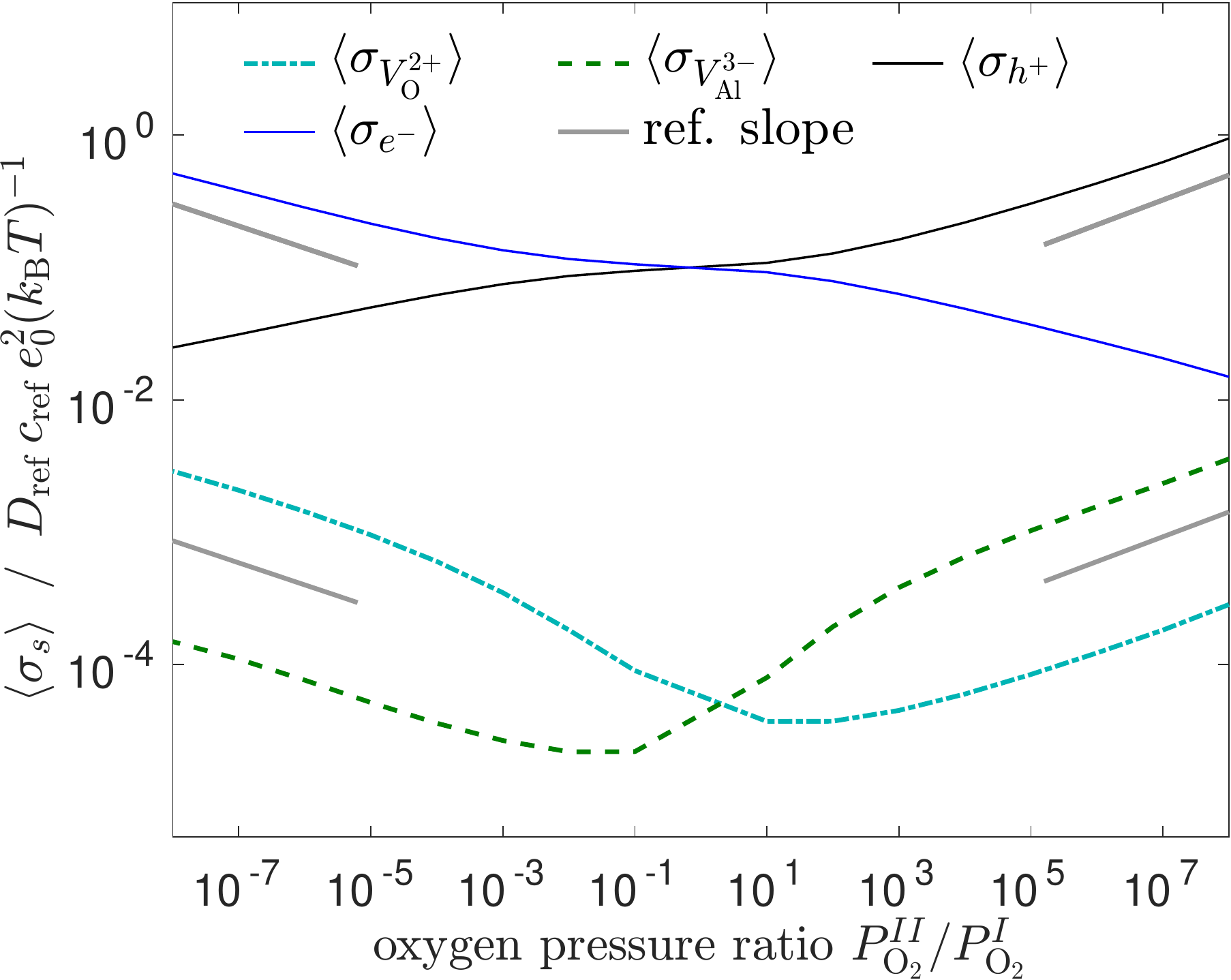}
  \end{center}
	\caption{Conductivities averaged over the grain boundary layer, $\langle \sigma_s \rangle$. 
		The reference lines are given by $\text{const.}\times\POtwo^{n_{\nu}}$.
	Scaled units with $e_0^2 \, \cref/\kBT \approx 1\,\text{C}^2\,\text{cm}^{-3}\,\text{J}^{-1}$.}
	\label{fig:cond-PO2}
\end{figure}

The plot for the average conductivities, see figure~\ref{fig:cond-PO2}, 
is equivalent to a Kr\"oger-Vink (Brouwer) diagram~\cite{Smyth2000} for the 
mobile species present, except that the average concentrations are scaled by the corresponding diffusion coefficients.
The averaged conductivities of the dominant point defect species, $\nu$, adhere to 
\begin{align} \label{eqn:sigma-n}
	\frac{\partial \, \text{ln} \langle \sigma_{\nu} \rangle }{ \partial \, \text{ln}\,\PII} = n_{\nu}.
\end{align}
The crossover in the dominant point defect species in the permeation rate 
and the average conductivity is observed at the same value of $\PII$.

\subsubsection{The Schottky equilibrium}

As discussed in section~\ref{sec:powlaw}, Schottky equilibrium and electron-hole equilibrium are required
conditions for the validity of the Wagner model and equation~\ref{eqn:wagner}. 
In this section we examine to what extent these equilibria are attained in our grain boundary calculations as the 
steady-state limit is approached.

\begin{figure}[H]
 \begin{center}
	 \includegraphics[width=6.5cm]{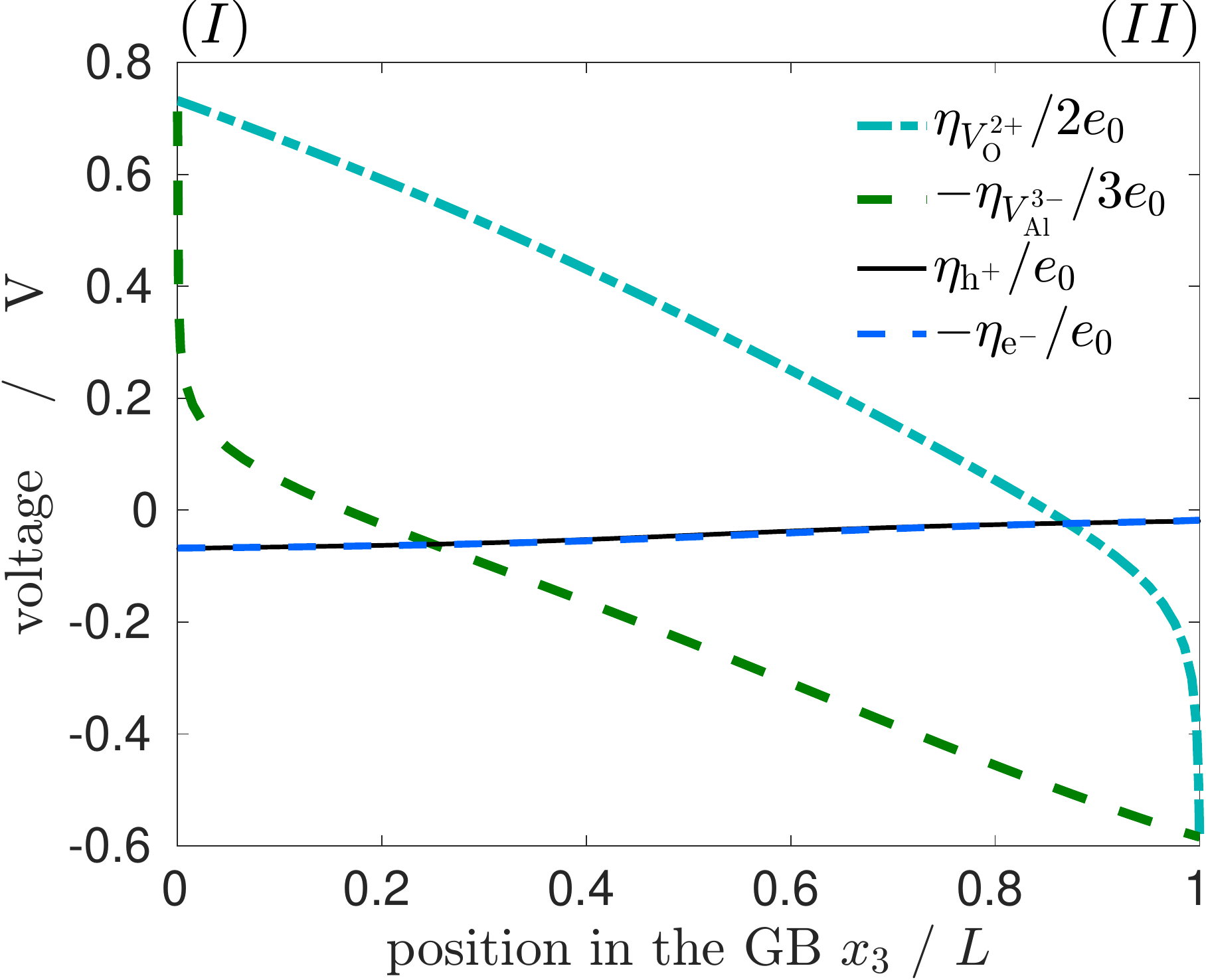}
  \end{center}
	\caption{The voltages, $\eta_s/z_se_0$, calculated from the ideal solution electrochemical potentials, 
	of the mobile species for fixed $x_1=\ahex/4$ in the grain boundary plane. 
Schottky equilibrium is not satisfied in the grain boundary since $-\eta_{\VAl}/3 \neq \eta_{\VO}/2$, but 
electrons and holes are in equilibrium, $-\eta_{\eprime} = \eta_{\hdot}$. 
Calculation for 
$\PII/\PI=10^{6}/10^{-9}$, $D_{\VO}=10^{-3} \,\Dref$, $D_{\VAl}=0.01\,\Dref$, 
and $D_{\eprime}=D_{\hdot}=1\,\Dref$, for other parameters see section~\ref{sec:init}.
The thermal voltage is given by $\kBT/e_0 = 0.16\,\text{V}$, 
and $V_{\eprime}=-(\eta_{\eprime}^{II}-\eta_{\eprime}^I)/e_0 = 50\,\text{mV}$. }
	\label{fig:eta_s}
\end{figure}

\begin{figure}[H]
 \begin{center}
	 \includegraphics[width=6.5cm]{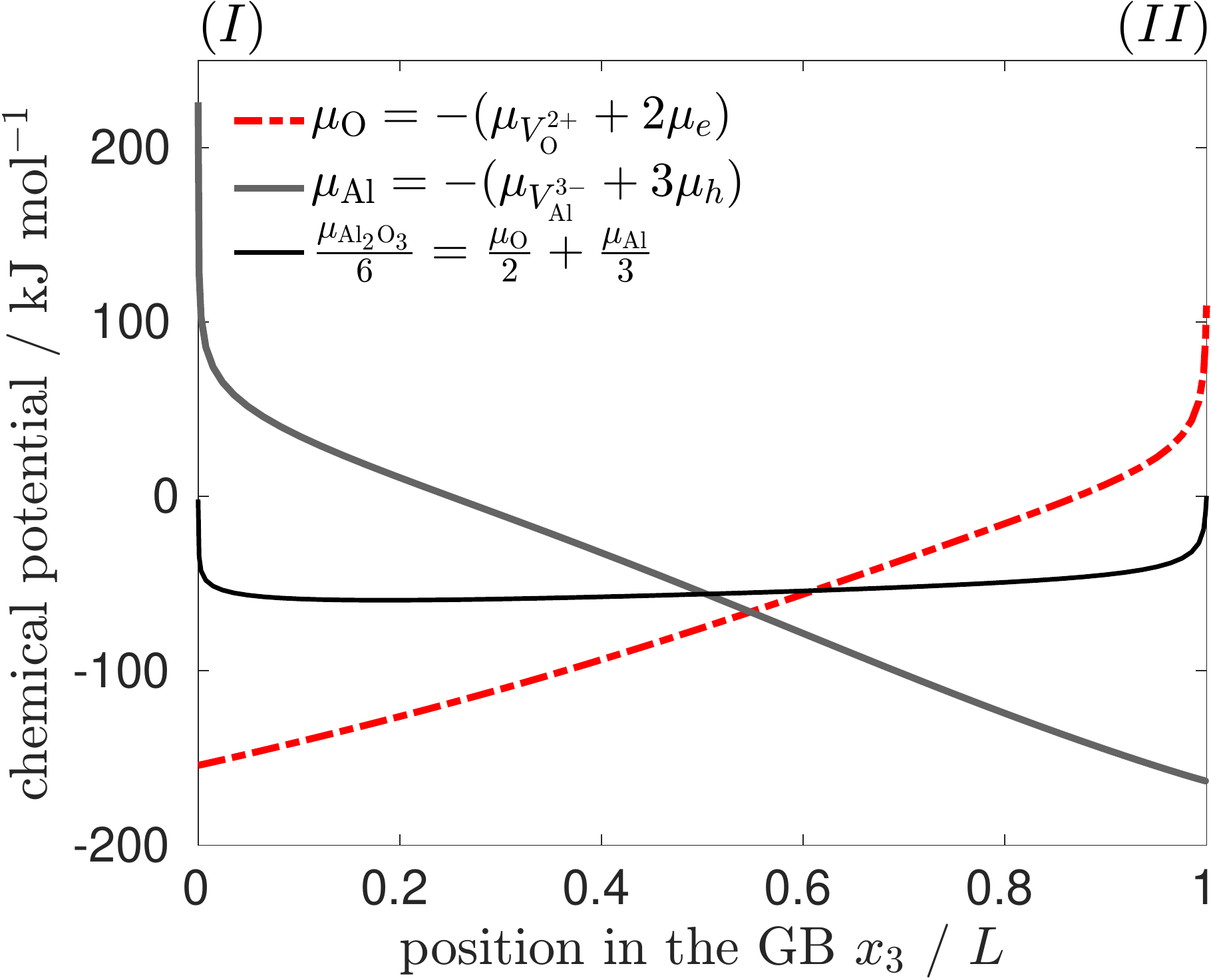}
  \end{center}
	\caption{Component aluminum and oxygen chemical potential, $\muAl$, and $\muO$, for fixed $x_1=\ahex/4$ in the 
	grain boundary plane. Schottky equilibrium holds at the surfaces where, $\mu_{\alumina}=0$, but
is not satisfied in the grain boundary. Same parameters are used as in figure~\ref{fig:eta_s}.}
	\label{fig:mu-comp}
\end{figure}

Figure~\ref{fig:eta_s} shows an example of the voltages, $\eta_s/z_se_0$, in the long time limit, the component 
chemical potentials are shown in figure~\ref{fig:mu-comp}. 
The calculations are performed for $\PI=\Pl$, and $\PII=\Ph$. 
In these calculations Schottky equilibrium does not hold, see figures~\ref{fig:eta_s}, and~\ref{fig:mu-comp}, 
except at the surfaces where the Schottky reaction is included in the equations.
The reaction rates at the surfaces are higher than the transport between the surfaces and the grain boundaries,
therefore Schottky equilibrium is attained at the surfaces.
The defect chemical potentials are calculated from the ideal dilute solution approximation, 
$\mu_s = \kBT \, \text{ln}(c_s/c_s(t_0))$, where $c_s(t_0)$ are the equilibrium concentrations.
The spatial variation of the electrochemical, and 
chemical potentials would be significantly different if internal Schottky equilibrium was imposed.

The component chemical potential distributions $\muAl(x_3)$, and $\muO(x_3)$ shown in figure~\ref{fig:mu-comp} 
are different from those shown in figure 8 of reference~\cite{Wada2011} for 
polycrystalline alumina membrane permeation experiments. 
Schottky equilibrium together with electron-hole equilibrium implies $3\muO + 2 \muAl=0$, 
which is not satisfied within the grain boundary, see figure~\ref{fig:mu-comp},
because $3\mu_{\VO} + 2\mu_{\VAl} \neq 0$.
The qualitative discrepancy in the chemical potential distributions between reference~\cite{Wada2011} 
and this treatment is likely due to the assumption of Schottky equilibrium in~\cite{Wada2011}.

A Schottky reaction term could be added to the transport equations~\ref{eqn:conti}, 
which would lead to Schottky equilibrium in the grain boundaries depending on the reaction and transport rates.
However, the internal Schottky reaction with formation and dissolution of oxide at the grain boundaries would 
induce stress and cause a non-trivial modification in the permeation rate.
The simulations discussed in this work are thought to correspond to the limit in which stress at the grain boundaries
prohibits the formation and dissolution of oxide internally, similarly to its effect in the bulk material.

\section{Discussion} \label{sec:sum}

The power laws for the oxygen permeation rate found experimentally in the limits of high and low applied oxygen pressures,
are confirmed in the calculations 
with the slab model in the steady-state limit. The experiments are performed on $0.25\,$mm thick polycrystalline 
alumina membranes while the geometry of the slab in the calculations
with $L=\ahex=1 \, \mu$m is chosen to resemble more closely 
the situation of planar films growing with a columnar grain structure.
The variation of the permeation rate, and the average conductivity
with the oxygen chemical potential are 
related to the power law exponent $n_{\nu}$, see figures~\ref{fig:PL} and~\ref{fig:cond-PO2}, 
and equations~\ref{eqn:P-n} and \ref{eqn:sigma-n}.
The power law exponent in turn depends 
on the stoichiometry of the quasi-chemical reactions at the surfaces, see equation~\ref{eqn:nnu}.
The applied pressure at which the transition between  
$p$-type and $n$-type ionic conductivity of the grain boundaries takes place depends on the 
ratio of the vacancy diffusion coefficients 
in the grain boundary, $D_{\VO}/D_{\VAl}$, 
given that $\VO$, and $\VAl$ are the mobile vacancy species.

The time-dependent calculations not only elucidate the initial transient behaviour but also help to clarify
the steady-state. 
Varying the applied oxygen partial pressure from $\Peq$ to $\Ph$ or $\Pl$ 
on one of the surfaces changes the rate of creation and annihilation
of vacancies, electrons and holes in stoichiometric proportions on the surface, 
the resulting chemical potential gradients between the surfaces along the grain boundaries
drive the transport of species through the slab. 
The resulting fluxes of the mobile species depend on their diffusion coefficients and 
will therefore not necessarily preserve the stoichiometric proportions of the surface populations of the species,
except when all species have the same diffusion coefficient.
The local charge density thereby becomes non-zero, and generates an electric field that retards the 
diffusion of the fastest dominant species, 
but enhances the diffusion of the slower dominant species which has an effective charge of opposite sign.

Three time scales 
involved in the dynamics of species concentration evolution are identified, $\tau_c$ (see equation~\ref{eqn:tauc}) 
characterizes the rate of charge build up on the surface and in the grain boundary, $\tau_d$ (see equation~\ref{eqn:taud})
characterizes the discharging of the grain boundary, and $\tau_s$ (see equation~\ref{eqn:taus}) 
characterizes the time to reach the steady-state at which $\nabla \cdot J_s \simeq 0$.

During the initial transient, $t \lesssim \tau_c$, 
the different diffusion coefficients lead to the accumulation of charge
in the grain boundary and on surface ($II$). 
If, for example in the $\Ph$ case, holes and aluminum vacancies are
generated on surface ($II$), the holes, which are assumed to be the faster species, $D_{\VAl} \ll D_{\hdot}$, 
diffuse away from the surface, 
generating positive charge density in the grain boundary, 
and surface ($II$) becomes negatively charged due to the aluminum vacancies.
Once the holes reach surface ($I$) it becomes positively charged, and
as the slower aluminum vacancies diffuse into the grain boundary it is discharged over a time $t\sim \tau_d$. 
In this example the electric field effectively enhances the aluminum 
vacancy diffusion, and retards the diffusion of the oxygen vacancies and holes.
In the steady-state limit the electrostatic potential difference 
between the surfaces, $\Delta \phi= \phi^{II}-\phi^I$, 
is on the order of the thermal voltage, $\kBT/e_0$, 
in the high and low pressure limits considered.

For the calculation shown in figure~\ref{fig:eta_s} with $\PII=\Ph$ and $\PI=\Pl$
the voltage between the surfaces 
observed in the simulations is $V_{\eprime}=-(\eta_{\eprime}^{II}-\eta_{\eprime}^I)/e_0 = 50\,\text{mV}$. 
$V_{\eprime}$ depends strongly on the species diffusion coefficients.

Physically, the mobile species distributions reconfigure to screen the electric field arising 
due to non-zero local charge density, and the Debye screening length (see equation~\ref{eqn:Debye}) 
characterizes the spatial extent of the 
variations in the local charge density. 
Due to screening effects the local charge density and electrostatic potential are challenging
to resolve numerically near the surfaces.
For spatial variations in the charge density within the 
grain boundary the two-dimensional equivalent of the reference screening length, see equation~\ref{eqn:lD2d}, 
is found to be appropriate. 
Apart from the initial transient the electrostatic potential in the simulations is dominated by the
contributions from the surface charges.
In the parameter regimes investigated 
the non-zero charge density in the grain boundary does not affect the defect fluxes significantly,
the 3D calculations can therefore be mapped onto a 1D model.
In the intermediate pressure regime the dominance of one vacancy species is less pronounced and in 
particular for $c_{\nu}(r,t) \sim c_{\nu}(t_0)$ in the steady-state the 
local charge neutrality approximation becomes invalid. 
For small species concentrations the Debye length gets larger,
and the electric field gets weaker. Both of the later facts are 
considered responsible for the failure of the analytic model in the intermediate pressure regime,
see figure~\ref{fig:PL}.

The simulations with the fixed oxygen chemical potential difference between
the surfaces reach a stationary non-equilibrium state with a constant rate of entropy production.
This means the populations of vacancies, electrons, and holes reach a dynamic equilibrium state, in which 
they are created and annihilated at the same rate and $\frac{\partial c_s}{\partial t} \simeq 0$, 
while the oxygen chemical potential difference sustains non vanishing fluxes between the surfaces.

\section{Conclusions}

A model for time-dependent grain boundary diffusion of ions and electrons through a film of polycrystalline oxide
has been constructed, in the form of a slab comprising hexagonal columnar grains. 
The long-range Coulomb interactions between the grain boundary planes affect the mass transfer dynamics through 
the slab significantly only during the initial transient. The electric field generated by the evolution of the charge
density influences the transport significantly; in the long time limit it is dominated by the surface charges.

Four mobile defects, charged aluminum and oxygen vacancies, electrons and holes, 
were considered, and simulations were made to compare with the behaviour observed in 
alumina oxygen permeation experiments.
The power laws for the permeation rate in the alumina membrane calculations 
depend on the stoichiometry of the quasi-chemical reactions at the membrane surfaces.

Work is in progress to extend the model; for example by including a Schottky reaction in the grain boundary we 
can couple the fluxes to the development of internal stress.

We have introduced a simplified, one-dimensional analytic model that employs the approximation of 
a single dominant defect, which appears from the 3D calculation to be justified. It does not assume internal 
Schottky equilibrium which is often assumed in Wagnerian treatments of oxidation.
The analytic model is shown to agree well with the simulation results under certain limiting conditions,
but fails in the intermediate oxygen partial pressure regime.

\section{Acknowledgements}
M.P.T was supported through a studentship in the Centre for Doctoral Training on Theory and Simulation of Materials
at Imperial College London funded by the Engineering and Physical Sciences Research Council under grant EP/G036888/1.
The authors would like to acknowledge the funding and technical support from BP 
through the BP International Centre for Advanced Materials (BP-ICAM) which made this research possible.
We are grateful for discussions with Profs.~Arthur Heuer, Matthew Foulkes and Brian Gleeson, and Dr.~Paul Tangney. 
The authors acknowledge support of the Thomas Young Centre under grant TYC-101.

\section*{References}
\bibliography{references.bib}

\appendix

\section{The system of reaction equations} \label{app:reac}
Applying the law of mass action to the reactions,~\ref{eqn:RO},
	\ref{eqn:RM}, \ref{eqn:RS}, and \ref{eqn:Rel} yields,
\begin{align}
	\mathcal{R}_{\text{O}} &= k_{\text{O},b} \left( K_{\text{O}} \, \POtwo^{1/2} \, \mathcal{C}_{\VO} -  \mathcal{C}_{\hdot}^2 \right)\\
	\mathcal{R}_{\text{Al}} 
	&= k_{\text{Al},b} \left(  K_{\text{Al}}\, \POtwo^{1/2}\, \mathcal{C}_{\eprime}^2 -\mathcal{C}_{\VAl}^{2/3}\right)\\
	\mathcal{R}_{\text{S}} &= -k_{\text{S},b} \left(\mathcal{C}_{\VAl}^{2/3} \, \mathcal{C}_{\VO} - K_{\text{S}}\right)\\
	\mathcal{R}_{\text{eh}} &= 	-k_{\text{eh},b} (\mathcal{C}_{\eprime} \, \mathcal{C}_{\hdot} - K_{\text{eh}}) 
\end{align}
and the coupled system of reactions at the alumina -- oxygen gas surfaces can be written as
\begin{subequations}\label{eqn:Rsys}
\begin{align} 
	\mathcal{R}_{\VO} &= -\mathcal{R}_{\text{O}} +  \mathcal{R}_{\text{S}} \hspace{1cm}
	\mathcal{R}_{\VAl} =  \mathcal{R}_{\text{Al}} + \frac{2}{3} \mathcal{R}_{\text{S}}\\
	\mathcal{R}_{\hdot} &= 2 \,\mathcal{R}_{\text{O}} + \mathcal{R}_{\text{eh}} \hspace{1cm}
	\mathcal{R}_{\eprime} = -3 \, \mathcal{R}_{\text{Al}} + \mathcal{R}_{\text{eh}}
\end{align}
\end{subequations}

\section{The analytic model} \label{app:mod-deriv}
The current density of the mobile species $s$ is expressed by
\begin{align}  \label{eqn:flux-app}
	&I_s = z_s e_0 J_s = -\sigma_s \left( \frac{\nabla \mu_s}{z_s e_0} -  \nabla \phi \right)  \\
	&\text{with} \hspace{1cm} \sigma_s = \frac{D_s c_s z_s^2 e_0^2}{\kBT}. 
\end{align}
We give here a derivation valid for the fluxes in the steady-state limit, 
assuming that one point defect species dominates and that there is no local excess of charge. 
Considering only the flux of the dominant vacancy species, $\nu\in\{\VO,\VAl\}$, 
and the charge compensating electronic species, $\nuo \in \{\hdot,\eprime \}$, 
their current densities in the steady state must balance at any point in the slab
\begin{align}
	I_{\nu} = - I_{\nuo}.
\end{align}
Inserting equation~\ref{eqn:flux-app} into this condition results in a constraint on the electric field,
\begin{align} \label{eqn:elec}
	-(\sigma_{\nu} + \sigma_{\nuo})\,\nabla \phi = \sigma_{\nu} \frac{\mu_{\nu}}{z_{\nu} e_0}+
	\sigma_{\nuo} \frac{\mu_{\nuo}}{z_{\nuo} e_0} 
\end{align}
and the current density, which can therefore be written as
\begin{align}
	I_{\nu} = \frac{\sigma_{\nu} \sigma_{\nuo}}{\sigma_{\nu} + \sigma_{\nuo}} 
	\, \frac{\nabla \mu_{\nu}+ |z_{\nu}| \nabla \mu_{\nuo}}{z_{\nu} e_0}.
\end{align}
In the steady-state $I_{\nu}$ does not depend on the position in the slab, 
hence upon integration over the instantaneous thickness of the slab, $L$,
\begin{align}\label{eqn:Inu-int}
	I_{\nu} L = 
	\int_{0}^{L} \frac{\sigma_{\nu} \sigma_{\nuo}}{\sigma_{\nu} + \sigma_{\nuo}}
	\frac{\nabla \mu_{\nu}+ |z_{\nu}| \nabla \mu_{\nuo}}{z_{\nu} e_0} dx.
\end{align}
The zero net current assumption is always expected to hold in the steady-state, and the 
above expression for the current density is expected to be a very good approximation.
The local charge neutrality approximation is less general and used to approximate the integral, it is given by,
\begin{align}\label{eqn:lcn}
	z_{\nu} c_{\nu} + z_{\nuo} c_{\nuo} = 0
\end{align}
and by using $|z_{\nuo}|=1$,
\begin{align}
\frac{\sigma_{\nu} \sigma_{\nuo}}{\sigma_{\nu} + \sigma_{\nuo}} = 
\frac{D_{\nu} D_{\nuo}}{D_{\nu}|z_{\nu}| + D_{\nuo}}  \frac{ c_{\nu} z_{\nu}^2 e_0^2}{\kBT	}.
\end{align}
Using the ideal solution approximation in addition to the local charge neutrality approximation,
\begin{align}
	d\mu_{\nu} &= \kBT d \left( \text{ln}\,c_{\nu} \right) = \kBT d \left( \text{ln} \, c_{\nuo} \right) = d\mu_{\nuo}
\end{align}
and hence
\begin{align}
	d\mu_{\nu} + |z_{\nu}| d\mu_{\nuo} &= (|z_{\nu}|+1) \, d\mu_{\nu}
\end{align}
the integral is given by,
\begin{align}
	I_{\nu} L = z_{\nu} e_0 \frac{D_{\nu} D_{\nuo}}{D_{\nu}|z_{\nu}| + D_{\nuo}} (|z_{\nu}|+1)
	\int_{0}^{L} \nabla c_{\nu} \, dx
\end{align}
and evaluated to yield the current density
\begin{subequations}
\begin{align}
	I_{\nu} &= z_{\nu} e_0\, D_{\nu}^{\text{eff}} \, \frac{\mathcal{C}_{\nu}^{II}-\mathcal{C}_{\nu}^I}{L},\\
	D_{\nu}^{\text{eff}} &= \frac{D_{\nu} D_{\nuo}}{D_{\nu}|z_{\nu}| + D_{\nuo}}(|z_{\nu}|+1). 
\end{align}
\end{subequations}

\end{document}